\documentclass[aps,nofootinbib,notitlepage,longbibliography,twocolumn,superscriptaddress,preprintnumber]{revtex4-1}

\usepackage{amsmath,amssymb,amsfonts}
\usepackage{physics}
\usepackage{slashed}
\usepackage{graphicx}
\usepackage{bm}
\usepackage{float}
\usepackage[T1]{fontenc}
\usepackage[utf8]{inputenc}
\usepackage{multirow}
\usepackage{gauss}
\usepackage[normalem]{ulem}
\usepackage[top=1.0in,bottom=1.0in,left=1.0in,right=1.0in]{geometry}
\usepackage{ragged2e}
\usepackage{tensor}
\usepackage{booktabs}
\usepackage{qcircuit}
\usepackage{fancyhdr}
\usepackage{lipsum}

\usepackage[colorlinks=true,linkcolor=blue,citecolor=blue,urlcolor=blue]{hyperref}

\usepackage{subcaption}
\usepackage{booktabs}
\usepackage{caption}
\usepackage{qcircuit}
\newcommand{\be}{\begin{equation}}
\newcommand{\ee}{\end{equation}}
\newcommand{\bea}{\begin{eqnarray}}
\newcommand{\eea}{\end{eqnarray}}
\newcommand{\ba}{\begin{eqnarray}}
\newcommand{\ea}{\end{eqnarray}}

\def\be{\begin{eqnarray}}
\def\ee{\end{eqnarray}}
\def\bea{\be}
\def\eea{\ee}

\def\roughly#1{\mathrel{\raise.3ex\hbox{$#1$\kern-.75em%
\lower1ex\hbox{$\sim$}}}}

\usepackage{lipsum}

\usepackage{fancyhdr}

\begin{document}
%\title{Role of anomalies in high energy QCD}
\title{Anomalies, Topology, and Hadron Structure in QCD}
\author{Ismail Zahed}
\email[]{ismail.zahed@stonybrook.edu}
\affiliation{Center for Nuclear Theory, Department of Physics and Astronomy,
Stony Brook University, Stony Brook, New York 11794–3800, USA}

%========================================
\begin{abstract}

Quantum Chromodynamics (QCD) provides a remarkable realization of how
quantum effects reshape the symmetries of a classical field theory.
The axial anomaly links chirality to gauge-field topology and
underlies the resolution of the $U(1)_A$ problem, while the trace
anomaly generates the intrinsic QCD scale through dimensional
transmutation and accounts for most of the mass of hadrons and hence
visible matter. Together, these quantum effects reveal the central
role of gluonic dynamics and vacuum structure in strong-interaction
physics.
In this review, we discuss the theoretical foundations and physical
consequences of anomalous symmetry breaking in QCD. We examine
anomalous Ward identities, the topological structure of gauge fields,
instantons, topological susceptibility, and the realization of chiral
and scale symmetries in the QCD vacuum. We review their role in the
generation of hadron masses, the $\eta'$ mass, and the resolution of
the $U(1)_A$ problem, and discuss their manifestations in polarized
deep inelastic scattering, nucleon spin structure, and modern studies
of hadron structure. Particular emphasis is placed on recent
developments connecting vacuum topology, the flavor-singlet axial
charge, topological screening, and the proton spin problem.
Our aim is to provide a unified perspective on how quantum anomalies
connect vacuum structure, hadron properties, and partonic observables,
bridging nonperturbative dynamics and perturbative QCD.

\end{abstract}

%\maketitle
%\tableofcontents

\maketitle
\tableofcontents 
\thispagestyle{fancy}
\fancyhead{}
%\fancyhead[R]{IQuS@UW-21-115, NT@UW-25-17}
\fancyfoot{}
\renewcommand{\headrulewidth}{0pt}

%========================================
\section{Introduction}
\label{sec:intro}
%========================================

Quantum Chromodynamics (QCD) is the non-Abelian gauge theory of the
strong interaction describing quarks and gluons through the color gauge
group $SU(3)$. Its dynamics determine the structure of hadrons and the
behavior of strongly interacting matter across a wide range of energy
scales. Although the QCD Lagrangian has a compact form,
\begin{equation}
\mathcal{L}_{\text{QCD}}
=
-\frac14 G_{\mu\nu}^a G^{a\mu\nu}
+
\sum_{f=1}^{N_f}
\bar{\psi}_f
\left(
i\gamma^\mu D_\mu - m_f
\right)
\psi_f ,
\label{eq:QCDlag}
\end{equation}
its physical consequences are highly nontrivial because of the
interplay between asymptotic freedom, confinement, spontaneous chiral
symmetry breaking, and vacuum topology
\cite{Gross:1973id,Politzer:1973fx}.
The gluon field strength tensor is
\begin{equation}
G_{\mu\nu}^a
=
\partial_\mu A_\nu^a
-
\partial_\nu A_\mu^a
+
g f^{abc} A_\mu^b A_\nu^c ,
\end{equation}
and the covariant derivative acting on quark fields is
\begin{equation}
D_\mu
=
\partial_\mu
-
igA_\mu^aT^a .
\end{equation}
The quark fields $\psi_f(x)$ carry flavor and color indices, while the
gluon fields $A_\mu^a(x)$ carry color,
with $T^a$ the generators of the color group $SU(3)$.

At short distances, asymptotic freedom permits a perturbative
description in terms of weakly interacting quarks and gluons. At
hadronic scales, however, QCD becomes strongly coupled and develops a
nonperturbative vacuum structure characterized by confinement,
condensates, and topological gauge-field fluctuations. Understanding
how these microscopic quark and gluon degrees of freedom generate the
observable properties of hadrons remains one of the central problems of
strong interaction physics.

A particularly important organizing principle in QCD is symmetry.
Neglecting light-quark masses, the classical QCD Lagrangian possesses
an approximate chiral symmetry,
\begin{equation}
SU(N_f)_L\times SU(N_f)_R ,
\end{equation}
together with classical scale invariance. These symmetries strongly
constrain the structure of the theory and its low-energy dynamics.

The physical vacuum, however, does not respect the full chiral symmetry
of the classical Lagrangian. The nonzero quark condensate
\begin{equation}
\langle0|\bar qq|0\rangle\neq0
\end{equation}
signals spontaneous chiral symmetry breaking and leads to the
appearance of light pseudoscalar mesons as pseudo-Goldstone bosons
\cite{Weinberg:1978kz,Gasser:1984gg}. Semiclassical approaches based on
instanton ensembles further suggest that topological gauge
configurations play an important role in generating the chiral
condensate and effective low-energy quark interactions
\cite{Diakonov:1985eg,Schafer:1996wv,Nowak:1996aj}.

In addition to spontaneous symmetry breaking, classical conservation
laws are modified by quantum anomalies. The singlet axial current
satisfies the anomalous divergence relation
\begin{equation}
\partial_\mu J_5^\mu
=
\frac{g^2N_f}{16\pi^2}
G_{\mu\nu}^a\tilde G^{a\mu\nu}
+
2\sum_f
m_f
\bar\psi_f i\gamma_5\psi_f ,
\label{eq:axialanomalyintro}
\end{equation}
where
\begin{equation}
J_5^\mu
=
\sum_f
\bar\psi_f\gamma^\mu\gamma_5\psi_f .
\label{eq:axialcurrentintro}
\end{equation}
which is the celebrated Adler-Bell-Jackiw anomaly
\cite{Adler:1969gk,Bell:1969ts}.
This relation links chirality directly to gauge-field topology through
the topological density
$G_{\mu\nu}^a\tilde G^{a\mu\nu}$ and plays a central role in the
resolution of the $U(1)_A$ problem, the structure of the QCD vacuum,
and the spin structure of the nucleon
\cite{Adler:1969gk,Bell:1969ts,Witten:1979vv,Veneziano:1979ec,Zahed:2022wae,Shuryak:2026pqt}.

Quantum effects also break classical scale invariance through the trace
anomaly,
\begin{equation}
T^\mu_{\ \mu}
=
\frac{\beta(g)}{2g}
G_{\mu\nu}^aG^{a\mu\nu}
+
\sum_f
m_f\bar\psi_f\psi_f ,
\label{eq:traceanomalyintro}
\end{equation}
which generates an intrinsic QCD mass scale through dimensional
transmutation. The trace anomaly is therefore directly connected to the
emergence of hadron masses and to modern studies of mass decomposition
and gravitational form factors
\cite{Ji:1995sv,Ji:2021mtz,Polyakov:2019lbq,Zahed:2021fxk}.

The operators appearing in the axial and trace anomalies indicate that
the nonperturbative QCD vacuum plays an essential dynamical role.
Gauge-field configurations with nontrivial topology, including
instantons, interpolate between vacuum sectors of different winding
number and induce chirality-changing processes
\cite{Belavin:1975fg,tHooft:1976snw,Diakonov:1995ea,Schafer:1996wv,Nowak:1996aj,Shuryak:2026pqt}. These effects contribute to the
topological susceptibility, the anomalously large $\eta'$ mass, and the
infrared structure of the singlet axial channel
\cite{Witten:1979vv,Veneziano:1979ec,Diakonov:1995ea,Schafer:1996wv,Nowak:1996aj,Shuryak:2026pqt}.

The consequences of anomalous dynamics extend far beyond the traditional
domains of low-energy hadron physics. In polarized deep inelastic
scattering, anomaly-induced operator mixing connects quark and gluon
helicity contributions and plays a central role in the interpretation
of nucleon spin observables
\cite{Jaffe:1989jz,Ji:1996ek,Leader:2013jra,Deur:2018roz}.
The flavor-singlet axial current links the proton spin problem to the
same topological fluctuations responsible for the resolution of the
$U(1)_A$ problem. As a result, measurements of polarized structure
functions probe not only partonic degrees of freedom but also the
infrared topological response of the QCD vacuum
\cite{Shore:1991dv,Shore:1992eu,Tarasov:2021yll,Tarasov:2025mvn}.

More generally, the operator-product expansion and perturbative QCD
evolution provide the framework connecting short-distance scattering
observables to nonperturbative matrix elements of quark and gluon
operators. The resulting interplay between perturbative and
nonperturbative physics is particularly evident in polarized parton
distributions, generalized parton distributions, transverse-momentum
dependent distributions, and Wigner phase-space distributions, which
collectively encode the spin, momentum, and spatial structure of the
nucleon
\cite{Ji:1996ek,Ji:1996nm,Belitsky:2005qn,Lorce:2011kd,
Leader:2013jra,Deur:2018roz}.

An equally important manifestation of anomalous dynamics is the
generation of hadron mass. Through the trace anomaly, quantum effects
break classical scale invariance and generate the intrinsic QCD scale
$\Lambda_{\rm QCD}$ via dimensional transmutation. The same gluonic
interactions responsible for confinement and vacuum structure therefore
also generate the dominant fraction of the visible mass of ordinary
matter. Recent theoretical and experimental developments have made it
possible to investigate this connection through gravitational form
factors, the energy-momentum tensor, generalized parton distributions,
and lattice-QCD studies of hadron mass decomposition
\cite{Ji:1995sv,Ji:2021mtz,Polyakov:2018zvc,Polyakov:2019lbq,
Zahed:2021fxk}.

Viewed broadly, anomalies provide a unique bridge between the ultraviolet
and infrared regimes of QCD. The axial anomaly links chirality to
topology, the trace anomaly links scale symmetry breaking to mass
generation, and both reveal the dynamical importance of the QCD vacuum.
From instantons and topological susceptibility to polarized scattering,
parton helicity distributions, and the internal structure of hadrons,
anomalous Ward identities provide a common language connecting seemingly
disparate phenomena across many orders of magnitude in energy and
distance scales.

This review focuses on the relation between anomalies, vacuum
structure, and hadronic observables in QCD. We begin with the symmetry
structure of the QCD Lagrangian and spontaneous chiral symmetry
breaking. We then discuss the axial anomaly, gauge-field topology, and
the role of instantons and topological susceptibility in the structure
of the QCD vacuum. The trace anomaly and the emergence of hadron masses
are subsequently examined, followed by applications to polarized deep
inelastic scattering, nucleon spin decomposition, anomaly-induced
operator mixing, and helicity dynamics in high-energy QCD.

Special emphasis is placed on recent developments linking vacuum
topology to the flavor-singlet axial charge, topological screening,
small-$x$ helicity evolution, and the modern understanding of proton
spin. Throughout, we emphasize how nonperturbative gluonic dynamics
manifest themselves in experimentally accessible observables, providing
a unified perspective on the role of anomalies in strong-interaction
physics.

Our goal is to provide a coherent account of how anomalous Ward
identities, vacuum topology, and gluonic dynamics connect phenomena
ranging from pseudoscalar meson physics and hadron mass generation to
polarized scattering, nucleon structure, and contemporary studies of
QCD at both low and high energies.

%========================================
\section{Chiral Symmetry in QCD}
\label{sec:chiralsymm}
%========================================

Symmetry principles play a central role in modern quantum field theory,
and Quantum Chromodynamics is no exception. Among the symmetries of QCD,
chiral symmetry occupies a particularly important place because it governs
many qualitative features of hadronic physics. The existence of light
pions, the pattern of low-energy hadronic interactions, and important
aspects of nucleon structure can all be traced to the realization of
chiral symmetry in the strong interaction. Historically, current algebra
and chiral symmetry constraints predated QCD itself, and many low-energy
results later understood from QCD were first derived as symmetry theorems.
Modern understanding combines current algebra, anomaly constraints,
effective field theory, lattice gauge theory, and nonperturbative vacuum
dynamics into a unified description of how hadronic structure emerges from
quark and gluon degrees of freedom
\cite{Weinberg:1978kz,Gasser:1984gg,Schafer:1996wv,Diakonov:2002fq,Nowak:1996aj}.

At the classical level, the approximate masslessness of the light quarks
implies that the QCD Lagrangian possesses an enlarged global symmetry
acting independently on left- and right-handed quark fields. This
symmetry underlies current algebra, low-energy theorems, and chiral
effective field theory. In the quantum theory, however, the realization
of chiral symmetry becomes more subtle. Part of the symmetry is
spontaneously broken by the QCD vacuum, producing pseudo-Goldstone
bosons, while another part is modified by anomalous quantum effects
associated with gauge-field topology. The interplay between chiral
symmetry breaking, confinement, and topological gauge configurations is
one of the defining nonperturbative features of QCD
\cite{Shuryak:1981ff,Diakonov:1985eg,Vainshtein:1981wh,Diakonov:1995ea,Schafer:1995pz,Nowak:1996aj}.

In this section we examine the chiral symmetry structure of the QCD
Lagrangian, derive the associated vector and axial currents, and develop
the framework needed later for anomalous Ward identities, vacuum
topology, and nucleon spin structure.

\subsection{Chiral decomposition of the quark fields}

The starting point is the QCD Lagrangian~\eqref{eq:QCDlag}.
To expose the chiral structure of the theory it is useful to separate
each quark field into components with definite chirality using the
projection operators
\begin{equation}
P_L = \frac{1-\gamma_5}{2},
\qquad
P_R = \frac{1+\gamma_5}{2},
\label{eq:chiral_projectors}
\end{equation}
which satisfy
\begin{equation}
P_L^2=P_L,
\qquad
P_R^2=P_R,
\qquad
P_LP_R=0,
\qquad
P_L+P_R=1 .
\label{eq:projector_algebra}
\end{equation}
The left- and right-handed components are
\begin{equation}
\psi_{fL}=P_L\psi_f,
\qquad
\psi_{fR}=P_R\psi_f .
\label{eq:chiraldecomposition}
\end{equation}
In the massless limit these correspond to states of definite helicity.
Although chirality and helicity are conceptually distinct, the two notions
coincide for massless particles. Since the light quark masses are small
compared with hadronic scales, this approximation is often physically
useful in QCD and underlies many partonic descriptions of high-energy
scattering~\cite{Ji:2002xn,Ji:2003yj}.
%,DeTeramond:2021jnn}.

The decomposition is especially natural because
\begin{equation}
\{\gamma^\mu,\gamma_5\}=0 ,
\label{eq:gamma5anticomm}
\end{equation}
which implies
\begin{equation}
\gamma^\mu P_L = P_R \gamma^\mu ,
\qquad
\gamma^\mu P_R = P_L \gamma^\mu .
\label{eq:gamma_projection}
\end{equation}
As a result, the Dirac operator exchanges chirality, while the kinetic
term propagates left- and right-handed fields independently after
projection.

\subsection{Global chiral symmetry}

With the chiral decomposition, the fermionic part of the QCD Lagrangian
becomes
\begin{widetext}
\begin{equation}
\sum_{f=1}^{N_f}
\bar{\psi}_f i\gamma^\mu D_\mu \psi_f
=
\sum_{f=1}^{N_f}
\bigg(
\bar{\psi}_{fL} i\gamma^\mu D_\mu \psi_{fL}
+
\bar{\psi}_{fR} i\gamma^\mu D_\mu \psi_{fR}
-
m_f(\bar{\psi}_{fL}\psi_{fR}+\bar{\psi}_{fR}\psi_{fL})
\bigg).
\label{eq:chirallagrangian}
\end{equation}
\end{widetext}
The kinetic terms propagate left- and right-handed components
independently, whereas the mass term couples them together. In the
massless limit the two sectors decouple, giving rise to chiral symmetry.
The fermionic action is then invariant under independent unitary
rotations of left- and right-handed quark fields,
\begin{equation}
\psi_L \rightarrow L \psi_L,
\qquad
\psi_R \rightarrow R \psi_R ,
\label{eq:chiralrotation}
\end{equation}
where
\begin{equation}
L,R\in U(N_f).
\end{equation}

The classical symmetry group of massless QCD is therefore
\begin{equation}
U(N_f)_L \times U(N_f)_R ,
\label{eq:UNf_symmetry}
\end{equation}
or equivalently
\begin{equation}
SU(N_f)_L \times SU(N_f)_R
\times U(1)_V \times U(1)_A .
\label{eq:fullchiralsymmetry}
\end{equation}
The $U(1)_V$ symmetry corresponds to baryon number conservation, while
$SU(N_f)_L$ and $SU(N_f)_R$ generate independent flavor rotations of
left- and right-handed quarks. The $U(1)_A$ symmetry corresponds to
the global chiral rotation
\begin{equation}
\psi_f \rightarrow e^{i\alpha\gamma_5}\psi_f ,
\label{eq:U1Arotation}
\end{equation}
under which
\begin{equation}
\psi_{fL} \rightarrow e^{-i\alpha}\psi_{fL},
\qquad
\psi_{fR} \rightarrow e^{+i\alpha}\psi_{fR}.
\label{eq:U1A_LR}
\end{equation}

At the classical level these symmetries are exact in the massless limit.
In the quantum theory, however, the singlet axial symmetry is broken by
the axial anomaly
\cite{Adler:1969gk,Bell:1969ts,tHooft:1976snw}, while the non-Abelian
chiral symmetry is spontaneously broken by the QCD vacuum, producing
Goldstone bosons.

The symmetry structure becomes especially transparent in terms of vector
and axial transformations,
\begin{equation}
V = \frac12(L+R),
\qquad
A = \frac12(L-R),
\label{eq:vectoraxial}
\end{equation}
where vector transformations rotate left- and right-handed fields in
the same direction, while axial transformations rotate them oppositely.

A central consequence of spontaneous chiral symmetry breaking is the
formation of the quark condensate,
\begin{equation}
\langle \bar{q} q \rangle
=
\sum_{f=1}^{N_f}
\langle \bar{\psi}_{fL}\psi_{fR}
+
\bar{\psi}_{fR}\psi_{fL}\rangle ,
\label{eq:qqbar_condensate}
\end{equation}
which acts as an order parameter linking the two chiral sectors.
Instanton-induced quark zero modes provide one microscopic mechanism for
this left-right mixing
\cite{Diakonov:1985eg,Shuryak:1981ff,Diakonov:1995ea,
Verbaarschot:1993pm,Leutwyler:1992yt,Banks:1979yr}.

The chiral structure of QCD therefore underlies the construction of
effective field theories, anomalous currents, spontaneous symmetry
breaking, and the interpretation of polarized high-energy scattering.

\subsection{Noether currents}

Every continuous symmetry generates a conserved current through
Noether's theorem. In QCD the corresponding vector and axial currents
play a central role in hadron structure, weak interactions, pion
dynamics, and polarized scattering
\cite{Weinberg:1978kz,Gasser:1984gg,Ji:1996ek,Nowak:1996aj,Deur:2018roz}.

Under the infinitesimal vector transformation
\begin{equation}
\psi_f
\rightarrow
\left(
e^{i\alpha^a T^a}
\right)_{ff'}
\psi_{f'} ,
\label{eq:vectortransformation}
\end{equation}
Noether's theorem gives the flavor vector current
\begin{equation}
J_V^{\mu,a}
=
\sum_{f,f'}
\bar{\psi}_f
\gamma^\mu
(T^a)_{ff'}
\psi_{f'} .
\label{eq:vectorcurrent}
\end{equation}
Its divergence satisfies
\begin{equation}
\partial_\mu J_V^{\mu,a}=0 .
\label{eq:vectorconservation}
\end{equation}
The singlet vector current,
\begin{equation}
J_V^\mu
=
\sum_{f=1}^{N_f}
\bar{\psi}_f\gamma^\mu\psi_f ,
\label{eq:baryoncurrent}
\end{equation}
corresponds to baryon number conservation and remains exact in QCD.

Under the infinitesimal axial transformation
\begin{equation}
\psi_f
\rightarrow
\left(
e^{i\alpha^a\gamma_5 T^a}
\right)_{ff'}
\psi_{f'} ,
\label{eq:axialtransformation}
\end{equation}
the corresponding flavor axial current is
\begin{equation}
J_A^{\mu,a}
=
\sum_{f,f'}
\bar{\psi}_f
\gamma^\mu\gamma_5
(T^a)_{ff'}
\psi_{f'} .
\label{eq:axialcurrent}
\end{equation}

Using the equations of motion,
\begin{equation}
\partial_\mu J_A^{\mu,a}
=
2i
\sum_{f,f'}
\bar{\psi}_f
(MT^a)_{ff'}
\gamma_5
\psi_{f'} ,
\label{eq:axialdivergence}
\end{equation}
where
\begin{equation}
M={\rm diag}(m_1,\dots,m_{N_f}) .
\end{equation}

Thus the non-singlet axial current is conserved in the massless limit.
The singlet axial current, however, is modified by the axial anomaly, as
discussed in the next section.
Vector currents encode exact or approximate flavor symmetries, whereas
axial currents are directly sensitive to spontaneous symmetry breaking
and vacuum structure. Matrix elements of axial currents therefore probe
the spin and chiral properties of hadrons and connect low-energy
phenomenology with polarized scattering observables and spin sum rules
\cite{Ji:1996ek,Nowak:1996aj,Jaffe:1989jz,Leader:2013jra,Deur:2018roz}.

\subsection{Goldstone bosons and their current couplings}

Spontaneous breaking of a continuous global symmetry implies the
existence of massless scalar excitations known as Goldstone bosons
\cite{Goldstone:1961eq}. In QCD the axial charges do not annihilate the
vacuum, and the axial current develops a one-particle pole associated
with the Goldstone boson. Thus the Goldstone bosons are required by the
realization of the symmetry itself.
The number of Goldstone bosons equals the number of broken generators.
For QCD with $N_f$ light flavors,
\(
N_G = N_f^2 - 1 .
\)

For two light flavors these are the three pions, while for three light
flavors the pseudoscalar octet includes the pions, kaons, and the
$\eta$ meson. Their defining coupling to the non-singlet axial current
is
\begin{equation}
\langle 0 | J_A^{\mu,a}(0) | \pi^b(p) \rangle
=
if_\pi p^\mu \delta^{ab}.
\label{eq:fpi_def}
\end{equation}
Here $J_A^{\mu,a}$ denotes the flavor non-singlet axial current and
$f_\pi$ is the pion decay constant. Equation~(\ref{eq:fpi_def})
expresses that the pion is created from the vacuum by the broken axial
current.
Taking the divergence gives
\begin{equation}
\partial_\mu J_A^{\mu,a}
\sim
f_\pi m_\pi^2 \pi^a ,
\label{eq:PCAC_goldstone}
\end{equation}
which vanishes in the exact chiral limit. This is the statement of the
partially conserved axial current (PCAC) relation.

The Goldstone nature of the pion explains why low-energy pion
interactions are weak. Since Goldstone bosons couple derivatively,
their amplitudes vanish at zero momentum in the chiral limit. This
property underlies soft-pion theorems and current algebra relations
\cite{Weinberg:1966kf,Adler:1965ga,Dashen:1969eg}.

The existence of Goldstone bosons therefore provides direct evidence
that the QCD vacuum is not invariant under the full chiral symmetry of
the classical Lagrangian. The pion sector encodes the infrared dynamics
of spontaneous symmetry breaking and forms the foundation of chiral
effective field theory.

\subsection{Explicit symmetry breaking and the GMOR relation}

In nature the quark masses are small but nonzero. As a result the
chiral symmetry of QCD is only approximate. The quark masses explicitly
break the symmetry and give the pions a small mass. The quantitative
relation between explicit symmetry breaking and the chiral condensate is
provided by the Gell-Mann-Oakes-Renner relation
\cite{GellMann:1968rz},
\begin{equation}
m_\pi^2 f_\pi^2
=
-(m_u+m_d)
\langle \bar{q} q \rangle
+
\mathcal{O}(m_q^2).
\label{eq:GMOR}
\end{equation}

Equation~(\ref{eq:GMOR}) shows that the pion mass vanishes linearly with
the light quark masses, while the coefficient is determined by the
chiral order parameter $\langle \bar{q} q \rangle$ and the pion decay
constant $f_\pi$.

The GMOR relation illustrates a central feature of hadron physics:
whereas the masses of most hadrons remain finite in the chiral limit
because of confinement and dynamical mass generation, the pion mass is
controlled directly by explicit chiral symmetry breaking.

The same condensate responsible for spontaneous chiral symmetry breaking
also controls the leading quark-mass dependence of the pseudoscalar
spectrum. The GMOR relation therefore serves both as a low-energy
theorem and as a probe of the QCD vacuum. Its corrections are organized
systematically in chiral perturbation theory, while its microscopic
origin can be studied through lattice QCD, instanton-based models, and
the spectral properties of the Dirac operator
\cite{Gasser:1984gg,Leutwyler:1992yt,Verbaarschot:1993pm,
Schafer:1996wv,Wittig:2020jtm,Diakonov:2002fq}.

In effective field theory the quark mass matrix acts as a spurion field
transforming under chiral symmetry, allowing systematic expansions of
hadronic observables in powers of quark masses and external momenta
\cite{Weinberg:1978kz,Gasser:1983yg,Gasser:1984gg}.

Modern lattice simulations have confirmed the GMOR relation with high
precision and provided quantitative determinations of the chiral
condensate and low-energy constants governing the chiral expansion
\cite{Aoki:2019cca,Wittig:2020jtm}.

\subsection{Effective description of low-energy QCD and the WZW term}

At energies well below the hadronic scale, the relevant degrees of
freedom are the Goldstone bosons rather than quarks and gluons. Their
interactions are described by chiral perturbation theory
\cite{Weinberg:1978kz,Gasser:1983yg,Gasser:1984gg},
which provides a systematic expansion in powers of momenta and quark
masses while preserving the symmetry structure of QCD.

For $N_f$ light flavors the Goldstone fields are collected into the
unitary matrix
\begin{equation}
U(x)
=
\exp
\left(
\frac{i\pi^a(x)T^a}{f_\pi}
\right),
\label{eq:Ufield}
\end{equation}
where the generators satisfy
\begin{equation}
{\rm Tr}(T^aT^b)=\frac12\delta^{ab}.
\label{eq:generator_normalization}
\end{equation}
with the chiral field transforming as
$SU(N_f)_L\times SU(N_f)_R$,
\begin{equation}
U \rightarrow LUR^\dagger .
\label{eq:Utransformation}
\end{equation}

The leading-order chiral Lagrangian is
\begin{equation}
\mathcal{L}_{\chi{\rm PT}}
=
\frac{f_\pi^2}{4}
{\rm Tr}
(\partial_\mu U^\dagger \partial^\mu U)
+
\frac{f_\pi^2}{4}
{\rm Tr}
(\chi U^\dagger + U\chi^\dagger),
\label{eq:chiral_lagrangian}
\end{equation}
where $\chi$ encodes explicit symmetry breaking by the quark mass
matrix.
The derivative structure of the kinetic term reflects the Goldstone
nature of the pion fields: their interactions vanish at zero momentum
in the exact chiral limit. This property underlies soft-pion theorems
and many predictions of current algebra.

Systematic corrections are organized through higher-derivative operators
and loop effects, yielding a controlled expansion for low-energy
hadronic amplitudes. Chiral perturbation theory has been applied
successfully to pion scattering, pion-nucleon interactions, kaon
physics, electroweak processes, and hadronic form factors
\cite{Gasser:1984gg,Scherer:2002tk}.

An essential refinement is that the effective theory must reproduce the
anomalous Ward identities of the underlying microscopic theory. The
classic example is the anomalous decay
$\pi^0\to\gamma\gamma$, whose resolution is the axial anomaly
\cite{Adler:1969gk,Bell:1969ts}. In the effective theory this
information is encoded in the Wess-Zumino-Witten term
\cite{Wess:1971yu,Witten:1983tw},
\begin{widetext}
\begin{equation}
S_{\mathrm{WZW}}[U]
=
-\frac{N_c}{240\pi^2}
\int_{M^5} d^5x\,
\epsilon^{ABCDE}\,
{\rm Tr}
\!\left(
U^{-1}\partial_A U\,
U^{-1}\partial_B U\,
U^{-1}\partial_C U\,
U^{-1}\partial_D U\,
U^{-1}\partial_E U
\right).
\label{eq:WZW}
\end{equation}
\end{widetext}
Here the five-dimensional manifold satisfies
\begin{equation}
\partial M^5 = M^4 ,
\end{equation}
with physical spacetime identified as the boundary manifold.
The anomalous variation of Eq.~(\ref{eq:WZW}) reproduces the chiral
anomaly of the underlying theory. Once external gauge fields are
introduced, it generates anomalous pseudoscalar couplings including
$\pi^0\to\gamma\gamma$.

The WZW term is topological in origin and reflects global properties of
the chiral field configuration space. Its coefficient is quantized and
proportional to the number of colors $N_c$, linking low-energy meson
dynamics directly to the microscopic color structure of QCD
\cite{Witten:1983tw}.

More generally, the WZW term demonstrates that effective field theories
must reproduce not only ordinary symmetries but also the anomaly
structure of the microscopic theory. This idea plays an important role
in modern discussions of anomaly matching, generalized symmetries, and
topological quantum field theory
\cite{tHooft:1979atm,Gaiotto:2014kfa}.

\subsection{'t~Hooft anomaly matching}
\label{subsec:tHooft_matching}

The statement of 't~Hooft anomaly matching is simple yet far reaching. If a
global symmetry $G$ of a quantum field theory is exact, then the
associated anomalies are invariant under renormalization group flow.
Consequently, the anomaly structure computed in the ultraviolet (UV)
must be reproduced in the infrared (IR), regardless of whether the IR
theory is confining, Higgsed, or described by Goldstone modes. This
constraint was formulated by 't~Hooft as a probe of possible infrared
phases
\cite{tHooft:1979atm}.

Anomalies constrain the infrared because they represent an obstruction
to gauging a global symmetry. Coupling the symmetry currents to
background gauge fields makes this explicit: if the UV theory produces
a non-invariance of the generating functional under background gauge
transformations, the IR theory must reproduce the same non-invariance
through massless degrees of freedom or topological terms. In this way,
anomalies encode information that survives confinement and remains
visible in long-distance physics.

For massless QCD, the classical global symmetry is
\begin{equation}
SU(N_f)_L\times SU(N_f)_R\times U(1)_V\times U(1)_A .\nonumber
\end{equation}
as we noted earlier.
Because $U(1)_A$ is broken by the axial anomaly
\cite{Adler:1969gk,Bell:1969ts}, the exact continuous symmetry relevant
for anomaly matching is
\begin{equation}
G_{\rm exact}
=
SU(N_f)_L\times SU(N_f)_R\times U(1)_V .\nonumber
\label{eq:Gexact}
\end{equation}
To identify  anomalies, one introduces background gauge fields
\begin{equation}
A_{L\mu}=A_{L\mu}^a T^a,
\qquad
A_{R\mu}=A_{R\mu}^a T^a,
\qquad
B_\mu ,
\end{equation}
corresponding respectively to
$SU(N_f)_L$,
$SU(N_f)_R$,
and
$U(1)_V$.
The UV quarks behave as Weyl fermions charged under these background
fields. The resulting triangle diagrams generate anomalies such as
$SU(N_f)_L^3$,
$SU(N_f)_R^3$,
and mixed
$SU(N_f)^2U(1)_V$
anomalies. Schematically,
\begin{equation}
D_\mu J_{L}^{\mu,a}
\propto
\frac{N_c}{24\pi^2}\,
d^{abc}\,
\epsilon^{\mu\nu\rho\sigma}
(F_L)^b_{\mu\nu}(F_L)^c_{\rho\sigma},
\label{eq:leftanomaly}
\end{equation}
where
\begin{equation}
d^{abc}
=
2\,{\rm Tr}
\!\left(
\{T^a,T^b\}T^c
\right)
\end{equation}
is the symmetric invariant tensor of $SU(N_f)$.

Because anomaly coefficients are renormalization-group invariants, the
infrared theory must reproduce them. A fully gapped chirally symmetric
phase would fail to do so and is therefore excluded. In QCD the observed
infrared realization is spontaneous chiral symmetry breaking,
\begin{equation}
SU(N_f)_L\times SU(N_f)_R
\rightarrow
SU(N_f)_V .
\label{eq:anomaly_matching_breaking}
\end{equation}
The infrared degrees of freedom are Goldstone bosons, which do not
generate perturbative triangle anomalies. Instead, the anomalies are
encoded in the Wess-Zumino-Witten term
\cite{Wess:1971yu,Witten:1983tw}, whose anomalous variation reproduces
the UV non-invariance of the generating functional. Its coefficient is
quantized and equal to $N_c$, exactly matching the UV anomaly
coefficients.

Anomaly matching therefore provides a direct demonstration that the
anomaly structure of the microscopic quark theory is preserved in the
low-energy dynamics of hadrons. More broadly, it illustrates that
anomalies are not erased by confinement but reorganized into infrared
degrees of freedom and topological interactions.

Modern developments involving generalized symmetries, anomaly inflow,
and topological quantum field theory have substantially broadened this
perspective~\cite{Gaiotto:2014kfa,Cordova:2022ruw}.

The anomaly matching viewpoint will reappear throughout this review. It
provides a conceptual bridge between ultraviolet operator identities and
infrared effective descriptions, clarifying why anomalous Ward
identities and topological structures remain visible in hadronic
observables.

\subsection{Connection to nucleon structure}

Chiral symmetry strongly constrains baryon structure, particularly
through the axial current. The nucleon matrix element of the isovector
axial current defines the axial charge $g_A$,
\begin{equation}
\langle p,s|J_A^{\mu,a}|p,s\rangle
=
2g_A\,s^\mu\,\delta^{a3},
\label{eq:gA_definition}
\end{equation}
where $|p,s\rangle$ denotes a nucleon state of momentum $p^\mu$ and
polarization $s^\mu$, normalized according to
\begin{equation}
s^2=-1,
\qquad
p\cdot s=0 .
\end{equation}
More generally,
\begin{widetext}
\begin{equation}
\langle p',s'|J_A^{\mu,a}|p,s\rangle
=
\bar{u}(p',s')
\left[
\gamma^\mu\gamma_5\, G_A(q^2)
+
q^\mu\gamma_5\, G_P(q^2)
\right]
\frac{\tau^a}{2}
u(p,s),
\label{eq:axial_formfactor_decomp}
\end{equation}
\end{widetext}
where
\begin{equation}
q^\mu=p'^\mu-p^\mu ,
\end{equation}
$G_A(q^2)$ is the axial form factor, and
$G_P(q^2)$ is the induced pseudoscalar form factor. The axial charge is
\begin{equation}
g_A = G_A(0).
\label{eq:gA_forward}
\end{equation}

Experimentally,
\begin{equation}
g_A \simeq 1.27 ,
\label{eq:gA_exp}
\end{equation}
making it one of the best measured quantities in low-energy QCD. It
governs neutron $\beta$ decay and provides a direct probe of nucleon
spin structure. Modern lattice calculations now determine $g_A$ with
increasing precision
\cite{Alexandrou:2020okk,Chang:2018uxx}.

The connection between axial structure and spontaneous chiral symmetry
breaking appears through the partially conserved axial current relation,
\begin{equation}
\partial_\mu J_A^{\mu,a}
=
f_\pi m_\pi^2 \pi^a + \cdots ,
\label{eq:PCAC}
\end{equation}
which shows that the axial current interpolates the pion field. This
leads directly to the Goldberger-Treiman relation,
\begin{equation}
g_A
=
\frac{f_\pi g_{\pi NN}}{M_N}
+
\mathcal{O}(m_\pi^2),
\label{eq:GoldbergerTreiman}
\end{equation}
where $M_N$ is the nucleon mass and $g_{\pi NN}$ the pion-nucleon
coupling. The relation illustrates how spontaneous chiral symmetry
breaking controls baryon properties through Goldstone-boson dynamics.

At the quark level the isovector axial current is
\begin{equation}
J_A^{\mu,a}
=
\sum_{f,f'}
\bar{\psi}_f
\gamma^\mu\gamma_5
\left(
\frac{\tau^a}{2}
\right)_{ff'}
\psi_{f'} .
\label{eq:isovector_axial_current}
\end{equation}
Its matrix elements probe the spin and helicity structure of the
nucleon and provide the starting point for polarized deep inelastic
scattering and spin sum rules. The singlet axial current will play a
particularly important role once anomaly effects are included in the
next section.

%========================================
\section{The Axial Anomaly in QCD}
\label{sec:axialanom}
%========================================

In the previous section we discussed the chiral symmetry structure of
the QCD Lagrangian and the associated conservation laws for vector and
axial currents in the classical theory. In the quantum theory,
however, the conservation of the flavor-singlet axial current cannot be
maintained simultaneously with gauge invariance. The resulting violation
of axial-current conservation is the axial anomaly, or
Adler-Bell-Jackiw anomaly
\cite{Adler:1969gk,Bell:1969ts}.

The axial anomaly provides one of the earliest and most profound
examples of a quantum symmetry breaking effect. It was discovered
independently by Adler and by Bell and Jackiw in their analyses of the
neutral-pion decay amplitude and the divergence of the axial current
\cite{Adler:1969gk,Bell:1969ts}. In
QCD it explains the large mass of the $\eta'$ meson, connects chiral
symmetry to gauge-field topology, and plays a central role in polarized
scattering and nucleon spin structure
\cite{Witten:1979vv,Veneziano:1979ec,tHooft:1976snw,
DiVecchia:1980yfw,Schafer:1996wv}.

In this section we derive the anomaly and develop its perturbative,
topological, and spectral interpretations.

\subsection{Axial current, anomaly, and topology}

Consider the flavor-singlet axial current defined earlier
\begin{equation}
J_5^\mu
=
\sum_{f=1}^{N_f}
\bar{\psi}_f\gamma^\mu\gamma_5\psi_f
=
\bar{\psi}\gamma^\mu\gamma_5\psi .\nonumber
\label{eq:singlet_axial_current}
\end{equation}
Using the on-shell Dirac equation,
\begin{equation}
(i\gamma^\mu D_\mu - M)\psi =0 ,
\label{eq:dirac_eq_anomaly}
\end{equation}
one finds the classical divergence relation
\begin{equation}
\partial_\mu J_5^\mu
=
2i\bar{\psi}M\gamma_5\psi
=
2i
\sum_{f=1}^{N_f}
m_f\bar{\psi}_f\gamma_5\psi_f .\nonumber
\label{eq:classical_axial_divergence}
\end{equation}
Thus finite quark masses explicitly break axial symmetry by mixing left-
and right-handed quark fields.

In the quantum theory the situation changes qualitatively. Ultraviolet
regularization modifies the divergence of the axial current in a way
that preserves gauge invariance but generates an additional
contribution from the fermion measure. The anomaly therefore reflects
the impossibility of preserving all classical symmetries
simultaneously after quantization.

\medskip

\noindent{\it Perturbative origin: triangle anomaly.}

The anomaly arises from the axial-vector-vector triangle diagram,
\begin{widetext}
\begin{equation}
T^{\mu\nu\rho}(p,q)
=
\sum_{f=1}^{N_f}
\int \frac{d^4k}{(2\pi)^4}
{\rm Tr}
\left[
\gamma^\mu\gamma_5
\frac{1}{\slashed{k}-m_f}
\gamma^\nu
\frac{1}{\slashed{k}+\slashed{p}-m_f}
\gamma^\rho
\frac{1}{\slashed{k}-\slashed{q}-m_f}
\right].
\label{eq:AVV_triangle}
\end{equation}
\end{widetext}
Because the integral is linearly divergent, its evaluation requires a
regularization scheme. Gauge invariance fixes the result uniquely and
leads to the anomalous Ward identity~Eq.(\ref{eq:axialanomalyintro}).
The details of the derivation are given in Appendix~\ref{app:anomaly_derivations}.

The divergence therefore contains both the explicit symmetry-breaking
mass term and a topological contribution generated by quantum effects.
The anomaly coefficient receives no higher-order perturbative
corrections, as guaranteed by the Adler-Bardeen theorem
\cite{Adler:1969er,Bardeen:1969md}.

\medskip

\noindent{\it Fujikawa path-integral derivation.}

Fujikawa showed that the anomaly originates from the noninvariance of
the fermionic path-integral measure under chiral transformations
\cite{Fujikawa:1979ay}. Under the infinitesimal
rotation
\begin{equation}
\psi
\rightarrow
e^{i\alpha(x)\gamma_5}\psi ,
\label{eq:fujikawa_rotation}
\end{equation}
the fermion measure transforms as
\begin{equation}
{\cal D}\bar{\psi}\,{\cal D}\psi
\rightarrow
J[\alpha]\,
{\cal D}\bar{\psi}\,{\cal D}\psi .
\label{eq:fujikawa_jacobian}
\end{equation}
After regulating the trace over fermionic eigenstates in a
gauge-invariant way, one recovers precisely the anomalous term
proportional to
$G_{\mu\nu}^a\tilde{G}^{a\mu\nu}$, as explained in Appendix~\ref{app:anomaly_derivations} and also next.
In this formulation the anomaly appears naturally as a property of the
functional measure itself.

\medskip

\noindent{\it Worldline representation.}

A geometrically intuitive derivation follows from the worldline
representation of the one-loop fermion effective action,
\begin{equation}
\Gamma[A]
=
- N_f
\int_0^\infty
\frac{dT}{T}\,
e^{-m^2T}
\int {\cal D}x\,{\cal D}\psi\,
e^{-S[x,\psi;A]},
\label{eq:worldline_effective_action}
\end{equation}
with worldline action
\begin{widetext}
\begin{equation}
S[x,\psi;A]
=
\int_0^T d\tau
\left(
\frac{\dot{x}^2}{4}
+
\frac12\psi_\mu\dot{\psi}^\mu
+
ig\dot{x}^\mu A_\mu^aT^a
-
ig\psi^\mu\psi^\nu G_{\mu\nu}^aT^a
\right).
\label{eq:worldline_action}
\end{equation}
\end{widetext}
Here \(x^\mu(\tau)\) describes the spacetime trajectory of the fermion
loop and the Grassmann variables \(\psi^\mu(\tau)\) encode its spin.
The ordinary fermion determinant corresponds to antiperiodic boundary
conditions for \(\psi^\mu\). The axial anomaly arises from the
non-invariance of the fermionic measure under an infinitesimal chiral
rotation,
%\[
%\psi\rightarrow e^{i\alpha\gamma_5}\psi,
%\qquad
%\bar\psi\rightarrow \bar\psi e^{i\alpha\gamma_5},
%\]
which yields the Fujikawa relation
\begin{equation}
\partial_\mu J^\mu_5
=
2im\,\bar\psi\gamma_5\psi
+
2\,{\rm tr}\,
\gamma_5
\langle x|
e^{-T\slashed D^2}
|x\rangle_{T\to0}.
\label{eq:fujikawa_worldline}
\end{equation}
The kernel in Eq.~(\ref{eq:fujikawa_worldline}) admits a worldline
representation,
\begin{equation}
{\rm tr}\,
\gamma_5
\langle x|
e^{-T\slashed D^2}
|x\rangle
=
\int_{x(0)=x(T)=x}
{\cal D}x\,{\cal D}\psi\,
e^{-S[x,\psi;A]},
\label{eq:gamma5_worldline_kernel}
\end{equation}
where the insertion of \(\gamma_5\) changes the Grassmann boundary
conditions from antiperiodic to periodic
\cite{Strassler:1992zr,Schubert:2001he}. The anomaly therefore probes
the periodic sector of the worldline path integral.

Periodic Grassmann fields possess four zero modes,
\begin{equation}
\psi^\mu(\tau)
=
\psi^\mu_0
+
\psi'^\mu(\tau),
\qquad
\int_0^T d\tau\,\psi'^\mu(\tau)=0 .
\label{eq:psi_zero_modes}
\end{equation}
A nonvanishing Grassmann integral requires saturation of all four
zero modes,
\begin{equation}
\int d^4\psi_0\,
\psi_0^\mu
\psi_0^\nu
\psi_0^\alpha
\psi_0^\beta
=
\epsilon^{\mu\nu\alpha\beta}.
\label{eq:grassmann_epsilon}
\end{equation}

The relevant interaction in the worldline action is
\begin{equation}
S_{\rm int}
=
-ig
\int_0^T d\tau\,
\psi^\mu\psi^\nu
G_{\mu\nu},
\label{eq:worldline_spin_field}
\end{equation}
and the leading contribution to the zero-mode sector arises from the
second-order term
\begin{equation}
S_{\rm int}^2
\propto
g^2
\left(
\psi^\mu\psi^\nu G_{\mu\nu}
\right)
\left(
\psi^\alpha\psi^\beta G_{\alpha\beta}
\right).
\label{eq:worldline_second_order}
\end{equation}
Integrating over the zero modes yields
\begin{equation}
\epsilon^{\mu\nu\alpha\beta}
G_{\mu\nu}^aG_{\alpha\beta}^a
=
2\,G_{\mu\nu}^a\widetilde G^{a\mu\nu},
\label{eq:worldline_zero_modes}
\end{equation}
which is precisely the topological.
In this form the anomaly emerges from the fermionic zero modes of the
periodic worldline sector. The resulting
\(\epsilon^{\mu\nu\alpha\beta}\) tensor is the worldline counterpart
of the index density appearing in the Atiyah-Singer theorem, making
the connection between spectral flow, zero modes, and the axial
anomaly particularly transparent.

\medskip

\noindent{\it Topological interpretation.}

The pseudoscalar gluonic operator
\begin{equation}
G_{\mu\nu}^a\tilde{G}^{a\mu\nu}
\label{eq:GdualG}
\end{equation}
is a total derivative and defines the topological charge density. Its
spacetime integral gives the winding number,
\begin{equation}
Q
=
\frac{g^2}{32\pi^2}
\int d^4x\,
G_{\mu\nu}^a\tilde{G}^{a\mu\nu}
\in \mathbb{Z}.
\label{eq:topological_charge}
\end{equation}

Gauge configurations with different values of $Q$ belong to distinct
topological sectors. Instantons provide semiclassical tunneling
solutions connecting these sectors
\cite{Belavin:1975fg,tHooft:1976snw,Callan:1976je,Schafer:1996wv,Nowak:1996aj}.
The anomaly therefore links chirality directly to vacuum topology:
axial charge can be exchanged with gauge-field topology in nontrivial
backgrounds.

\medskip

\noindent{\it Index theorem and spectral flow.}

The Atiyah-Singer index theorem relates the topological charge to the
spectrum of the Euclidean Dirac operator
\cite{Atiyah:1968mp},
\begin{equation}
n_+ - n_- = N_f Q,
\label{eq:index_theorem}
\end{equation}
where $n_\pm$ are the numbers of right- and left-handed fermion zero
modes.

In time-dependent gauge backgrounds, eigenvalues of the Dirac operator
flow through zero as the gauge field evolves between sectors of
different topology. The resulting change in axial charge is
\begin{equation}
\Delta Q_5 = 2N_f Q,
\label{eq:deltaQ5}
\end{equation}
which is the integrated form of the anomaly equation.

The perturbative triangle diagram, Fujikawa measure analysis,
worldline formulation, and spectral-flow interpretation therefore
provide complementary descriptions of the same underlying phenomenon:
the axial anomaly is an intrinsic feature of gauge theories linking
ultraviolet regularization, fermion spectra, and gauge-field topology.

\subsection{Relation to chiral symmetry breaking}

The axial anomaly plays a central role in determining how chiral
symmetry is realized in QCD. As discussed in
Sec.~\ref{sec:chiralsymm}, the classical massless theory possesses the
global symmetry shown in Eq.~(\ref{eq:fullchiralsymmetry}).

If all axial symmetries were spontaneously broken, one would expect
$N_f^2$ Goldstone bosons. Experimentally, however, only
$N_f^2-1$ light pseudoscalar mesons are observed. The missing singlet
Goldstone boson is removed by the axial anomaly.

For the singlet axial current, the divergence relation is given by
Eq.~(\ref{eq:axialanomalyintro}). The anomalous term survives even in the
chiral limit and therefore destroys conservation of the $U(1)_A$
current at the quantum level. Unlike the non-singlet chiral
symmetries, $U(1)_A$ is not an exact symmetry of QCD and cannot
generate a true Goldstone boson.

%The realized symmetry-breaking pattern is therefore
%\begin{equation}
%SU(N_f)_L\times SU(N_f)_R
%\rightarrow
%SU(N_f)_V ,
%\label{eq:SSB_anomaly_section}
%\end{equation}
%which produces the observed pseudoscalar octet for $N_f=3$.

Physically, the anomaly couples quark chirality to gauge-field
topology. Instanton transitions between vacuum sectors with different
winding number change the axial charge according to
Eq.~(\ref{eq:deltaQ5}). The vacuum therefore acts as a source and sink
of axial charge, lifting the would-be singlet Goldstone mode through
topological fluctuations~\cite{Zahed:2022wae} (and references therein). We will elaborate further on this below.

An alternative understanding of this mechanism was also developed by Witten
and Veneziano in the large-$N_c$ limit, where the $\eta'$ mass is
related to the topological susceptibility of pure Yang-Mills theory
\cite{Witten:1979vv,Veneziano:1979ec}.

The anomaly also affects polarized scattering and nucleon spin
observables. In the singlet channel, anomalous operator mixing couples
quark and gluon contributions under renormalization-group evolution
\cite{Jaffe:1989jz,Ji:1996ek,Leader:2013jra}. Modern discussions of
gluon helicity, proton spin decomposition, and topological
contributions to axial charge are therefore deeply connected to the
anomaly structure developed here
\cite{Qian:2015wyq,Zahed:2022wae,Shuryak:2021fsu}.

%=======================================

\subsection{Anomaly and nucleon spin}

The flavor-singlet axial current plays a central role in the spin
structure of the nucleon. Its forward matrix element defines the
singlet axial charge,
\begin{equation}
\langle P,S|J_5^\mu|P,S\rangle
=
2M\,S^\mu\,g_A^{(0)},
\label{eq:gA0_def}
\end{equation}
which measures the net helicity carried by quarks and antiquarks in
the proton.

Unlike the non-singlet axial currents, the singlet current is not
conserved even in the chiral limit because of the anomaly
(\ref{eq:axialanomalyintro}). Consequently, the interpretation of
$g_A^{(0)}$ differs from that of ordinary conserved charges. The
anomaly induces a mixing between quark and gluon helicity operators,
linking polarized deep inelastic scattering directly to gluonic
degrees of freedom
\cite{Jaffe:1989jz,Altarelli:1988nr,Bass:2004xa}.

The resulting decomposition of the proton spin involves quark spin,
gluon spin, and orbital angular momentum contributions,
\begin{equation}
\frac12
=
\frac12\Delta\Sigma
+
\Delta G
+
L_q
+
L_g ,
\label{eq:spinsumrule_intro}
\end{equation}
where the precise separation of the various terms depends on the
operator definition and factorization scheme. Modern formulations based
on gauge-invariant decompositions, generalized parton distributions,
and phase-space distributions will be discussed in
Sec.~\ref{sec:spin}.

The anomalous divergence equation also suggests a deeper connection
between nucleon spin and vacuum topology. Since the axial anomaly
couples the singlet current to the topological density
$G\widetilde G$, topological gauge-field fluctuations may contribute
to the flavor-singlet axial charge. This possibility connects the
proton spin problem to the same vacuum topology that underlies the
$U(1)_A$ problem. These issues are discussed in detail in
Sec.~\ref{sec:spin} below.

%=============================

%====================================================

\subsection{Physical interpretation}

The axial anomaly provides one of the clearest demonstrations that a
classical symmetry need not survive quantization. In QCD the anomaly
emerges from the simultaneous requirements of gauge invariance,
ultraviolet regularization, and the spectral structure of the Dirac
operator in nontrivial gauge backgrounds.

The perturbative triangle diagram, Fujikawa path-integral formulation,
worldline representation, and spectral-flow picture discussed earlier
provide mathematically equivalent descriptions of the same phenomenon.
Together they show that the anomaly is not merely a perturbative
artifact but a global property of gauge theories with chiral fermions.

An important feature is that the anomaly is exact. The coefficient of
the anomalous divergence receives no perturbative corrections beyond one
loop, as guaranteed by the Adler-Bardeen theorem
\cite{Adler:1969er,Bardeen:1969md}. This robustness makes the anomaly a
powerful constraint on both ultraviolet and infrared descriptions of
QCD.

The anomaly links several central themes of QCD into a common framework:
chiral symmetry breaking, vacuum topology, instantons, topological
susceptibility, the $\eta'$ mass, and nucleon spin structure all emerge
from the interplay between fermions and non-Abelian gauge fields.

Modern developments involving anomaly matching, generalized symmetries,
and topological response theories have further emphasized the role of
anomalies as exact probes of strongly coupled dynamics
\cite{tHooft:1979atm,Gaiotto:2014kfa,Cordova:2022ruw}.

%%%%%%%%%%%%%%%%%%%%%%%%%%%%%%%%%%%%%%%%%%%%%%%%%%%%%%%%%%%%%%%%%%%%%%%%%%%%%%
\subsection{Anomalies in hard exclusive processes}
\label{subsec:dvcs_anomaly}
%%%%%%%%%%%%%%%%%%%%%%%%%%%%%%%%%%%%%%%%%%%%%%%%%%%%%%%%%%%%%%%%%%%%%%%%%%%%%%

While the axial anomaly was originally established through the
triangle diagram and its connection to vacuum topology, recent work has
shown that anomalous Ward identities also play an important role in
hard exclusive processes. In particular, deeply virtual Compton
scattering (DVCS) provides a unique framework in which the axial and
trace anomalies enter through local operator matrix elements appearing
in the operator-product expansion and QCD factorization. This extends
the role of anomalies beyond low-energy effective theories and vacuum
structure into observables probing the three-dimensional quark and
gluon structure of hadrons.

Within the QCD factorization framework, the axial and trace anomalies
have recently been analyzed systematically in DVCS
\cite{Bhattacharya:2022laz,Bhattacharya:2023jdn}. These studies show
that the anomalous Ward identities constrain the short-distance
coefficient functions together with the matrix elements of local
operators entering generalized parton distributions and
energy-momentum tensor form factors. They also clarify how the
trace anomaly contributes to the decomposition of the Compton
amplitude and how these contributions are consistently incorporated
within perturbative QCD beyond leading order.

Related perturbative analyses have revisited matrix elements of the
axial and vector currents, emphasizing the role of ultraviolet and
infrared regularization, operator renormalization, and scheme
dependence in preserving the anomaly relations. These studies provide a
complementary perturbative perspective on the anomaly, making explicit
how anomalous Ward identities emerge in renormalized current matrix
elements entering high-energy scattering amplitudes.

These developments complement the viewpoint adopted throughout this
review. Here the emphasis is placed on the nonperturbative origin of
the axial and trace anomalies through vacuum topology, instantons,
topological susceptibility, and spectral flow. The modern factorization
approach demonstrates that the same anomaly operators continue to
govern hard exclusive reactions through generalized parton
distributions and energy-momentum tensor matrix elements. Together,
these developments provide a unified picture in which anomalies connect
vacuum structure, hadron tomography, and perturbative QCD.

%====================================================

\subsection{Transition to vacuum structure}

The appearance of the topological density in Euclidean signature
\[
G_{\mu\nu}^a\tilde{G}^a_{\mu\nu}
\]
in the anomaly equation signals that axial symmetry is intimately tied
to the topology of gauge fields.

As discussed above, this operator is a total derivative whose spacetime
integral measures the winding number of the gauge configuration. The
QCD vacuum therefore contains distinct topological sectors connected by
nonperturbative tunneling processes mediated by instantons and related
gauge configurations
\cite{Belavin:1975fg,tHooft:1976snw,Schafer:1996wv}.

The anomaly thus provides the conceptual bridge between chiral symmetry
and vacuum topology. Through this connection, topological fluctuations
influence the breaking of $U(1)_A$, the generation of the $\eta'$
mass, vacuum condensates, and the distribution of spin and axial charge
inside hadrons.

These issues motivate a more detailed examination of the QCD vacuum,
which we now turn to in the next section.

%========================================
\section{The Structure of the QCD Vacuum}
\label{sec:vacuum}
%========================================

The vacuum of Quantum Chromodynamics is qualitatively different from the
vacuum of perturbative field theory. In QCD, strong gluonic
self-interactions generate a highly nontrivial vacuum characterized by
topological fluctuations, condensates, and long-range correlations.
The vacuum therefore acts as an active dynamical medium whose structure
determines many observable properties of hadrons.

As discussed in Sec.~\ref{sec:axialanom}, the axial anomaly relates the
divergence of the singlet axial current to the topological density
appearing in Eq.~(\ref{eq:axialanomalyintro}). Gauge configurations with
nonzero winding number therefore play a dynamical role in the quantum
theory and contribute directly to spontaneous chiral symmetry breaking,
the $\eta'$ mass, and aspects of hadron spin structure
\cite{tHooft:1976snw,Witten:1979vv,Veneziano:1979ec,
Schafer:1996wv,Shuryak:1981ff}.

More generally, the properties of the QCD vacuum are encoded in
correlation functions of gauge-invariant operators. These correlations
determine vacuum condensates, topological susceptibility, and the
emergence of the confinement scale $\Lambda_{\rm QCD}$. Modern
developments in lattice gauge theory, semiclassical methods,
Dyson-Schwinger approaches, and holographic models have all reinforced
the importance of nonperturbative vacuum dynamics in strong interaction
physics
\cite{Diakonov:2002fq,Wittig:2020jtm,Roberts:2021nhw,
Shuryak:1993ee}.

%====================================================

\subsection{Gauge field configurations and topology}

A defining feature of non-Abelian gauge theories is the existence of
topologically distinct gauge-field configurations. Consider the
Euclidean Yang-Mills action
\begin{equation}
S_{\rm YM}
=
\frac14
\int d^4x,
G_{\mu\nu}^a G_{\mu\nu}^a ,
\label{eq:YM_action}
\end{equation}
with Euclidean field strength \(G_{\mu\nu}^a.\)
Finite-action configurations satisfy
\begin{equation}
G_{\mu\nu}^a \rightarrow 0
\qquad
(|x|\rightarrow\infty),
\label{eq:finite_action_condition}
\end{equation}
so that asymptotically the gauge field approaches a pure gauge,
\begin{equation}
A_\mu
\rightarrow
\frac{i}{g}U^{-1}\partial_\mu U .
\label{eq:pure_gauge_asymptotic}
\end{equation}

Compactifying Euclidean spacetime at infinity identifies the boundary
with \(S^3\), and the gauge transformation defines a mapping
\begin{equation}
U:S^3\rightarrow SU(N_c).
\label{eq:topological_map}
\end{equation}
These mappings are classified by
\begin{equation}
\pi_3(SU(N_c))
=
\mathbb Z ,
\label{eq:homotopy_group}
\end{equation}
implying inequivalent topological sectors labeled by an integer winding
number.
The associated topological invariant is the Pontryagin index already
introduced in Eq.~(\ref{eq:topological_charge}),
where the Euclidean dual field strength is
\begin{equation}
\widetilde G_{\mu\nu}^a
=
\frac12
\epsilon_{\mu\nu\rho\sigma}
G_{\rho\sigma}^a ,
\qquad
\epsilon_{1234}=+1 .
\label{eq:dual_tensor_vacuum}
\end{equation}
For self-dual and anti-self-dual configurations,
\begin{equation}
G_{\mu\nu}^a
=
\pm
\widetilde G_{\mu\nu}^a ,
\label{eq:selfduality_topology}
\end{equation}
the action and topological charge are directly related. These
configurations correspond to instantons and anti-instantons and play a
central role in the nonperturbative dynamics of QCD.
The topological density is a total derivative,
\begin{equation}
G_{\mu\nu}^a
\widetilde G_{\mu\nu}^a
=
\partial_\mu K_\mu ,
\label{eq:GdualG_total_derivative}
\end{equation}
with \(K_\mu\) the Chern-Simons current. Consequently, the
topological charge depends only on the asymptotic structure of the
gauge field and is invariant under smooth deformations of the gauge
configuration.

Configurations with different winding numbers correspond to distinct
vacua that cannot be continuously connected through perturbative gauge
transformations. Quantum tunneling between these sectors therefore
produces intrinsically nonperturbative effects invisible in ordinary
perturbation theory. The corresponding tunneling trajectories are the
Euclidean instantons discussed below.

Throughout this discussion the gauge configurations are understood in
Euclidean spacetime, where instantons appear as finite-action saddle
points of the Yang-Mills action. Physical Minkowski observables such
as hadronic matrix elements, the trace anomaly, or the vacuum energy
density are obtained only after analytic continuation from the
underlying Euclidean theory.

%====================================================

\subsection{Instantons and tunneling between vacua}

The existence of distinct topological sectors implies that the QCD
vacuum is not unique. Classical vacuum configurations correspond to
pure gauges with different winding numbers,
\begin{equation}
|n\rangle ,
\qquad
n\in\mathbb Z ,
\label{eq:topological_vacua}
\end{equation}
which cannot be continuously deformed into one another by small gauge
transformations.

Quantum mechanically, transitions between these vacua occur through
tunneling processes mediated by instantons. Instantons are finite-action
solutions of the Euclidean Yang-Mills equations satisfying the
self-duality condition Eq.~(\ref{eq:selfduality_topology}).
Substituting this relation into the Yang-Mills action yields
\begin{equation}
S_{\rm inst}
=
\frac{8\pi^2}{g^2}|Q|,
\label{eq:instanton_action}
\end{equation}
so the tunneling amplitude behaves as
\begin{equation}
\mathcal A_{\rm tunnel}
\sim
\exp\!\left(
-\frac{8\pi^2}{g^2}
\right).
\label{eq:tunneling_amplitude}
\end{equation}
The exponential dependence on $1/g^2$ demonstrates explicitly that
instanton effects are intrinsically nonperturbative
\cite{Belavin:1975fg,tHooft:1976snw,Callan:1977gz}.

Instantons provide the microscopic realization of the axial anomaly.
Using the index theorem already discussed in
Eq.~(\ref{eq:index_theorem}),
an instanton with $Q=1$ changes the axial charge according to
Eq.~(\ref{eq:deltaQ5}), thereby inducing chirality-changing processes
in the vacuum.

Physically, instantons may be viewed as localized tunneling events that
rearrange the chirality structure of the Dirac sea. In the semiclassical
vacuum picture developed by Shuryak, Diakonov, and Petrov, the QCD
vacuum can be modeled as an interacting ensemble of instantons and
anti-instantons whose collective dynamics generate nontrivial infrared
physics
\cite{Shuryak:1981ff,DYAKONOV1984259,Schafer:1995pz,
Diakonov:1995ea}.

Instanton-induced interactions contribute to flavor mixing, hadronic
correlation functions, constituent quark masses, and spontaneous chiral
symmetry breaking through the collectivization of fermion zero modes
\cite{Diakonov:1985eg,Diakonov:2002fq,Schafer:1996wv}.

%====================================================

\subsection{The $\theta$ vacuum}

Because the classical vacua $|n\rangle$ are topologically distinct, the
physical ground state of QCD is the coherent superposition
\begin{equation}
|\theta\rangle
=
\sum_n e^{in\theta}|n\rangle .
\label{eq:theta_vacuum}
\end{equation}
The parameter $\theta$ introduces the additional term
\begin{equation}
\mathcal L_\theta
=
\theta\,
\frac{g^2}{32\pi^2}
G_{\mu\nu}^a\tilde G_{a\mu\nu}.
\label{eq:theta_term}
\end{equation}
Although the topological density is a total derivative, the $\theta$
term has physical consequences because of the nontrivial topology of
gauge configurations. Under parity and time reversal,
\begin{equation}
G_{\mu\nu}^a\tilde G_{a\mu\nu}
\rightarrow
-
G_{\mu\nu}^a\tilde G_{a\mu\nu},
\label{eq:CPodd_theta}
\end{equation}
so the $\theta$ term violates CP symmetry.

Experimental limits on the neutron electric dipole moment imply
\begin{equation}
|\theta|
\lesssim
10^{-10},
\label{eq:theta_bound}
\end{equation}
leading to the strong CP problem
\cite{Crewther:1979pi,Baker:2006ts}.
The best-known resolution is the Peccei-Quinn mechanism, in which a
dynamical axion field relaxes the effective $\theta$ parameter to zero
\cite{Peccei:1977hh,Weinberg:1977ma,Wilczek:1977pj}. The axion emerges
as a pseudo-Goldstone boson associated with a spontaneously broken
anomalous chiral symmetry and couples directly to the topological
density appearing in Eq.~(\ref{eq:theta_term}).
The resulting phenomenology connects QCD vacuum topology with cosmology
and dark matter physics.

Recent studies have provided a more direct connection between the
$\theta$ vacuum and measurable aspects of hadron structure.
Within the instanton description of the QCD vacuum, a finite
$\theta$ angle induces CP-odd quark interactions that generate
electric dipole moments and form factors for hadrons.
In particular, the nucleon electric dipole form factor was shown
to emerge from the same topological pseudoparticle ensemble that
governs the $\theta$ dependence of the vacuum energy and the
topological susceptibility~\cite{Liu:2025kuc}. For small $\theta$,
the induced CP-odd form factor is proportional to the topological
charge fluctuations of the vacuum and is related to the anomalous
chirality-changing interactions generated by instanton zero modes.
These results provide a concrete illustration of how the topology
of the QCD vacuum can be encoded in experimentally accessible
observables probing the internal structure of the nucleon.

%====================================================
\subsection{Topological susceptibility}

An important quantity characterizing topological fluctuations in the QCD
vacuum is the topological susceptibility 
%introduced previously in
%Eq.~(\ref{eq:topological_susceptibility}), where the topological charge
%density is defined in Eq.~(\ref{eq:topological_density})Equivalently,
\begin{equation}
\chi_{\rm top}
=
\left.
\frac{\partial^2E(\theta)}
{\partial\theta^2}
\right|_{\theta=0},
\label{eq:chi_top_theta}
\end{equation}
%so $\chi_{\rm top}$ 
which measures the response of the vacuum energy to
topological deformations. It therefore provides a direct quantitative
probe of topological charge fluctuations in the QCD vacuum.

%As discussed in connection with the Witten-Veneziano relation,
%$\chi_{\rm top}$ controls the anomalous contribution to the $\eta'$
%mass. 
In the chiral limit with massless quarks, the susceptibility is
suppressed because fermion zero modes screen topological fluctuations,
whereas in pure Yang-Mills theory it remains finite
\cite{Leutwyler:1992yt,DiVecchia:1980yfw}. These features are well reproduced in the QCD instanton vacuum~\cite{Zahed:2022wae,Liu:2024rdm} (and references therein).

Modern lattice gauge theory calculations have provided increasingly
precise determinations of topological susceptibility and the
distribution of topological charge in the QCD vacuum
\cite{Luscher:2009eq,Luscher:2010iy,Wittig:2020jtm}. Gradient-flow
methods and improved lattice operators have also enabled direct
visualization of topological structures such as instantons and vortices
in gauge configurations
\cite{Leinweber:1999cw,Biddle:2020eec}.

The connection between the anomalously large $\eta'$ mass and the
topological susceptibility of pure Yang-Mills theory emerges most
naturally in the large-$N_c$ limit through the Witten-Veneziano
mechanism \cite{Witten:1979vv,Veneziano:1979ec}, which we now explain.

\subsection{Connection with the $\eta'$ mass}

The most direct phenomenological manifestation of the axial anomaly is
the large mass of the $\eta'$ meson. In the absence of the anomaly, the
$\eta'$ would appear as the ninth pseudo-Goldstone boson associated
with spontaneous breaking of $U(1)_A$ and would therefore be
parametrically light. Instead,
\begin{equation}
m_{\eta'} \simeq 958~{\rm MeV},
\label{eq:etaprime_mass}
\end{equation}
far larger than the pion and kaon masses.
A quantitative relation between the anomaly and the $\eta'$ mass
emerges in the large-$N_c$ limit through the Witten-Veneziano
mechanism
\cite{Witten:1979vv,Veneziano:1979ec}. The relevant quantity is the
topological susceptibility in
Eq.~(\ref{eq:chi_top_theta}). %where the topological charge density is
%defined in Eq.~(\ref{eq:topological_density}).

The singlet anomaly relation, Eq.~(\ref{eq:axialanomalyintro}),
shows that the pseudoscalar singlet channel couples directly to
topological gauge fluctuations. Saturating the corresponding
correlation function with the lowest singlet pseudoscalar state yields
the Witten-Veneziano relation,
\begin{equation}
m_{\eta'}^2
=
\frac{2N_f}{f_\pi^2}\,
\chi_{\rm top}
+
\mathcal{O}\!\left(\frac{1}{N_c^2}\right).
\label{eq:WittenVeneziano}
\end{equation}

Equation~(\ref{eq:WittenVeneziano}) shows explicitly that the $\eta'$
mass is generated primarily by gluonic topology rather than by explicit
quark masses. Using the large-$N_c$ scaling relations
\begin{equation}
\chi_{\rm top}\sim \mathcal{O}(1),
\qquad
f_\pi^2\sim \mathcal{O}(N_c),
\label{eq:largeNc_scaling}
\end{equation}
one obtains
\begin{equation}
m_{\eta'}^2
\sim
\mathcal{O}\!\left(\frac{1}{N_c}\right),
\label{eq:eta_scaling}
\end{equation}
showing that the anomaly contribution disappears in the strict
large-$N_c$ limit.

The $\eta'$ therefore provides one of the clearest experimental
manifestations of vacuum topology in QCD. Lattice calculations of the
topological susceptibility now reproduce the observed $\eta'$ mass with
increasing precision, providing strong support for the role of the topological fluctuations permeating the QCD vacuum~\cite{Zahed:2022wae,Liu:2024rdm,Shuryak:2026pqt} (and references therein), and also supporting  the Witten-Veneziano result in quenched QCD
\cite{Alles:1996nm,DelDebbio:2004ns,Cichy:2015jra}.

%====================================================

\subsection{Vacuum condensates and the instanton liquid}

A useful microscopic picture of the QCD vacuum is provided by the
instanton liquid model (ILM), in which the vacuum is treated as an
ensemble of instantons and anti-instantons with characteristic size
$\bar\rho$ and density $n$. Phenomenologically, Shuryak noted long ago that~\cite{Shuryak:1981ff}
\begin{equation}
\bar\rho
\sim
0.3~{\rm fm},
\qquad
n
\sim
1~{\rm fm}^{-4},
\label{eq:ILM_parameters}
\end{equation}
corresponding to a semi-dilute medium of topological fluctuations.

Within this framework, spontaneous chiral symmetry breaking arises
naturally from the zero modes of the Dirac operator in instanton
backgrounds. For a single instanton,
\begin{equation}
i\slashed D\,\psi_0 = 0 ,
\label{eq:Dirac_zero_mode}
\end{equation}
with a localized chiral zero mode $\psi_0$~\cite{tHooft:1976snw}. In an ensemble of
instantons and anti-instantons, overlap between these zero modes
generates a finite density of near-zero Dirac eigenvalues~\cite{Diakonov:1995ea}.

The resulting quark condensate is related to the spectral density
through the Banks-Casher relation,
\begin{equation}
\langle\bar\psi\psi\rangle
=
-\pi\rho(0),
\label{eq:Banks_Casher}
\end{equation}
showing that spontaneous chiral symmetry breaking is tied directly to
the infrared structure of the Dirac spectrum
\cite{Banks:1979yr}. In fact there are infinitely many Banks-Casher-like relations, all of which are captured by random matrix theory~\cite{Verbaarschot:1993pm}. If Eq.~(\ref{eq:Banks_Casher}) is interpreted as the quark chiral conductivity at the quark line with zero virtuality  (the analogue of the Fermi surface in a conducting metal), then the additional relations capture the fluctuations of the chiral conductance at the mesoscopic scale.

Integrating out fermions in the instanton background generates the
effective multi-fermion interaction first derived by 't~Hooft~\cite{tHooft:1976snw},
\begin{equation}
\mathcal L_{\rm tHooft}
\sim
\prod_{f=1}^{N_f}
(\bar\psi_{fR}\psi_{fL})
+
{\rm h.c.},
\label{eq:tHooft_interaction}
\end{equation}
which breaks $U(1)_A$ while preserving the non-singlet chiral symmetry.
This interaction provides a microscopic realization of topology-induced
chirality-changing processes and contributes to the formation of the
quark condensate.

The same framework also generates the gluon condensate,
\begin{equation}
\left\langle
G_{\mu\nu}^aG^{a\mu\nu}
\right\rangle
\sim
\frac{32\pi^2}{g^2}n ,
\label{eq:gluon_condensate}
\end{equation}
indicating that both quark and gluon condensates are controlled by the
same underlying topological dynamics.

These condensates enter directly into the trace anomaly relation,
Eq.~(\ref{eq:traceanomalyintro}), 
demonstrating that nonperturbative gluonic fluctuations dominate the vacuum energy density~\cite{Liu:2024rdm}. Vacuum condensates therefore encode the dynamical
breaking of scale invariance and connect the microscopic QCD vacuum to
hadronic observables
\cite{Shifman:1978bx,Zahed:2021fxk,Liu:2024rdm,Shuryak:2026pqt}, as we will elaborate further below.

%====================================================

%========================================
\section{Conformal Symmetry and the Trace Anomaly}
\label{sec:traceanom}
%========================================

In addition to chiral symmetry, classical QCD with massless quarks also
possesses scale invariance. At the classical level the QCD Lagrangian
contains no intrinsic dimensionful parameter and is therefore invariant
under global rescalings of spacetime coordinates. Quantum effects,
however, fundamentally modify this picture. Renormalization introduces a
dynamical scale into the theory, and the resulting violation of scale
invariance is encoded in the trace anomaly.

The trace anomaly plays a role analogous to that of the axial anomaly
discussed in Sec.~\ref{sec:axialanom}. In the axial case, quantization
destroys conservation of the singlet axial current through the
topological density appearing in Eq.~(\ref{eq:axialanomalyintro}), whereas in
the conformal case quantization destroys scale invariance through the
scalar gluonic operator \(G_{\mu\nu}^aG^{a\mu\nu}.\)

The anomaly therefore provides the operator-level explanation for
dimensional transmutation, the emergence of $\Lambda_{\rm QCD}$, and
ultimately the origin of most hadronic masses. Unlike masses generated
through the Higgs mechanism, the dominant part of the nucleon mass
arises dynamically from strongly interacting gluonic fields and vacuum
fluctuations
\cite{Ji:1995sv,Yang:2018nqn,Ji:2021mtz,Roberts:2021nhw,Schafer:1996wv,Nowak:1996aj,Diakonov:1989un}.

%====================================================

\subsection{Classical scale symmetry}

In the chiral limit, the QCD Lagrangian 
Eq.~(\ref{eq:QCDlag}) is scale free
\begin{equation}
\mathcal L_{\rm QCD}
\rightarrow
-\frac14
G_{\mu\nu}^aG^{a\mu\nu}
+
\sum_{f=1}^{N_f}
\bar\psi_f i\gamma^\mu D_\mu\psi_f .
\label{eq:QCD_massless_lagrangian}
\end{equation}
Since the gauge coupling is dimensionless in four spacetime dimensions,
the action is invariant under global dilatations,
\begin{equation}
x^\mu \rightarrow \lambda x^\mu ,
\label{eq:dilatation_transform}
\end{equation}
provided the fields transform according to their canonical dimensions,
\begin{equation}
\psi_f(x)
\rightarrow
\lambda^{-3/2}\psi_f(\lambda x),
\qquad
A_\mu(x)
\rightarrow
\lambda^{-1}A_\mu(\lambda x).
\label{eq:scale_field_transform}
\end{equation}
The corresponding Noether current is the dilatation current,
\begin{equation}
D^\mu
=
x_\nu T^{\mu\nu},
\label{eq:dilatation_current}
\end{equation}
whose divergence satisfies
\begin{equation}
\partial_\mu D^\mu
=
T^\mu_{\ \mu}.
\label{eq:dilatation_divergence}
\end{equation}
Classical scale invariance therefore implies
\begin{equation}
T^\mu_{\ \mu}=0 .
\label{eq:classical_tracelessness}
\end{equation}

More generally, conformal invariance enlarges the symmetry group to the
full conformal group $SO(4,2)$, including dilatations and special
conformal transformations. In QCD, however, this symmetry is broken
quantum mechanically by renormalization effects. The resulting
dynamically generated infrared scale governs hadron masses,
confinement, and nonperturbative vacuum structure
\cite{Coleman:1970je,Callan:1970ze,Collins:1976yq}.

At short distances, asymptotic freedom causes the running coupling to
become small, and QCD approximately approaches a scale-invariant
theory. In this regime logarithmic scaling violations replace exact
Bjorken scaling, leading to the renormalization-group evolution of
parton distributions and correlation functions
\cite{Gross:1973id,Politzer:1973fx}.

%====================================================

\subsection{Renormalization and breaking of scale invariance}

The quantum theory requires ultraviolet regularization and
renormalization. Introducing a renormalization scale $\mu$ immediately
breaks classical scale invariance because physical quantities now depend
on the running coupling,
\begin{equation}
\mu\frac{dg}{d\mu}
=
\beta(g).
\label{eq:beta_definition}
\end{equation}
At one loop,
\begin{equation}
\beta(g)
=
-\frac{g^3}{16\pi^2}
\left(
\frac{11}{3}N_c
-
\frac{2}{3}N_f
\right)
+\mathcal O(g^5),
\label{eq:beta_oneloop}
\end{equation}
which is negative for
\begin{equation}
N_f < \frac{11}{2}N_c .
\label{eq:AF_condition}
\end{equation}
This negative beta function is the origin of asymptotic freedom: the
strong interaction becomes weak at short distances but strong at large
distances
\cite{Gross:1973id,Politzer:1973fx}.

The breaking of scale invariance follows directly from the
renormalization group. The renormalized generating functional satisfies
\begin{equation}
\mu\frac{dW}{d\mu}=0,
\label{eq:RG_W}
\end{equation}
which implies
\begin{equation}
\left(
\mu\frac{\partial}{\partial\mu}
+
\beta(g)\frac{\partial}{\partial g}
\right)W
=
0 .
\label{eq:CallanSymanzik}
\end{equation}
Scale transformations therefore become equivalent to changes in the
renormalization scale. Since the coupling runs, scale invariance no
longer survives quantization.

This phenomenon parallels the axial anomaly discussed earlier. In both
cases a classical symmetry fails after quantization because ultraviolet
regularization introduces additional structure into the theory. The
resulting anomaly is encoded in an exact operator identity relating the
divergence of a symmetry current to local gauge-invariant operators.

More generally, renormalization implies that coupling constants acquire
scale dependence through quantum fluctuations. In QCD the antiscreening
contribution from gluon self-interactions dominates over fermionic
screening, producing the negative beta function characteristic of
non-Abelian gauge theories
\cite{Gross:1973ju,Wilczek:1973jp}.

%====================================================

\subsection{The trace anomaly}

Under an infinitesimal Weyl transformation,
\begin{equation}
x^\mu\to e^\sigma x^\mu ,
\qquad
\mu\to e^{-\sigma}\mu ,
\end{equation}
the variation of \(W\) is generated by the trace of the energy-momentum tensor,
\begin{equation}
\delta_\sigma W
=
-\sigma\int d^4x\, T^\mu_{\ \mu}.
\end{equation}
But the same transformation changes the coupling by
\begin{equation}
\delta_\sigma g=-\sigma\,\mu\frac{dg}{d\mu}
=-\sigma\,\beta(g).
\end{equation}
so that
\begin{equation}
\delta_\sigma W
=
\int d^4x\,
\frac{\partial{\cal L}}{\partial g}\,\delta_\sigma g
=
-\sigma\int d^4x\,
\beta(g)\frac{\partial{\cal L}}{\partial g}.
\end{equation}
Now note that the variation of \(W\) is generated by the trace of the energy-momentum tensor,
\begin{equation}
\delta_\sigma W
=
-\sigma\int d^4x\, T^\mu_{\ \mu}.
\end{equation}
Equating the two expressions gives
\begin{equation}
T^\mu_{\ \mu}
=
\beta(g)\frac{\partial{\cal L}}{\partial g}.
\end{equation}

To evaluate the derivative, it is convenient to rescale the gauge field according to
\(
gA_\mu \rightarrow A_\mu ,
\)
so that the Yang-Mills Lagrangian takes the form
\begin{equation}
\mathcal L_g
=
-\frac{1}{4g^2}\,
\widehat G^a_{\mu\nu}\widehat G^{a\mu\nu},
\end{equation}, hence
\begin{equation}
T^\mu_{\ \mu}
=
\frac{\beta(g)}{2g}
G^a_{\mu\nu}G^{a\mu\nu}.
\end{equation}
Including explicit quark masses yields the full operator identity
\begin{equation}
T^\mu_{\ \mu}
=
\frac{\beta(g)}{2g}
G_{\mu\nu}^aG^{a\mu\nu}
+
\sum_{f=1}^{N_f}
m_f(1+\gamma_m)
\bar\psi_f\psi_f ,
\label{eq:full_trace_anomaly}
\end{equation}
where $\gamma_m$ is the anomalous dimension of the quark mass operator.

In the chiral limit, the entire trace is generated dynamically by the
gluonic operator.
The physical content is that quantum fluctuations make the coupling scale dependent. The introduction of the renormalization scale breaks classical scale invariance, and the resulting trace anomaly provides the operator realization of dimensional transmutation and the emergence of the dynamical scale $\Lambda_{\rm QCD}$.

The gluonic operator mixes with scalar quark operators under
renormalization, and the precise separation between quark and gluon
contributions depends on the renormalization scheme and scale
\cite{Collins:1976yq,Nielsen:1977sy}. These operator identities play an
important role in modern decompositions of hadron mass and momentum~\cite{Zahed:1994qh,Shuryak:2026pqt} (and references therein).

Because the energy-momentum tensor couples directly to the spacetime
metric, the trace anomaly also governs the response of QCD to curved
backgrounds and external gravitational fields
\cite{Duff:1993wm,Brown:1976wc}. In modern language, the trace anomaly
may therefore be viewed as the breaking of local Weyl symmetry by
quantum effects.
\subsection{Dimensional transmutation and the origin of mass}

The trace anomaly provides the field-theoretic realization of
dimensional transmutation. Solving the renormalization-group equation
introduces the dynamically generated scale
\begin{equation}
\Lambda_{\rm QCD}
=
\mu
\exp\!\left[
-
\int^g
\frac{dg'}{\beta(g')}
\right],
\label{eq:LambdaQCD}
\end{equation}
which replaces the dimensionless coupling by a physical mass scale.
Low-energy hadronic quantities are then set by
$\Lambda_{\rm QCD}$,
\begin{equation}
M_{\rm hadron}
\sim
\Lambda_{\rm QCD},
\label{eq:hadron_mass_scale}
\end{equation}
up to corrections from explicit quark masses. Although the classical
theory contains no intrinsic scale, quantum fluctuations dynamically
generate one through renormalization-group evolution.

The relation between hadron masses and the trace anomaly follows from
matrix elements of the energy-momentum tensor,
\begin{equation}
\langle P|
T^\mu_{\ \mu}
|P\rangle
=
2M_P^2 ,
\label{eq:Tmunu_hadron}
\end{equation}
showing that hadron masses arise predominantly from gluonic quantum
fluctuations rather than from bare quark masses.
For the proton,
\begin{equation}
M_p \simeq 938~{\rm MeV},
\label{eq:proton_mass}
\end{equation}
only a small fraction originates from the light $u$ and $d$ quark
masses. Most of the mass is generated dynamically through confinement,
gluon self-interactions, and the trace anomaly
\cite{Ji:1995sv,Yang:2018nqn,Ji:2021mtz,Hackett:2023rif}.

Current analyses based on lattice QCD, gravitational form factors, and
energy-momentum tensor matrix elements increasingly support this
picture of emergent hadron mass generation
\cite{Ji:1994av,Ji:1995sv,Lorce:2021xku}.

%====================================================

\subsection{Vacuum structure and gluon condensates}

The trace anomaly connects directly to the nonperturbative vacuum
structure discussed in Sec.~\ref{sec:vacuum}. Taking the vacuum
expectation value of the anomaly equation,
Eq.~(\ref{eq:full_trace_anomaly}) in the chiral limit, gives
\begin{equation}
\langle0|
T^\mu_{\ \mu}
|0\rangle
=
\frac{\beta(g)}{2g}
\langle0|
G_{\mu\nu}^aG^{a\mu\nu}
|0\rangle .
\label{eq:vacuum_trace}
\end{equation}
The vacuum energy density is therefore controlled by the gluon
condensate,
\[
\left\langle
\frac{\alpha_s}{\pi}
G_{\mu\nu}^aG^{a\mu\nu}
\right\rangle ,
\label{eq:gluon_condensate_trace}
\]
which characterizes long-range nonperturbative gluonic fluctuations.

The gluon condensate plays a central role in QCD sum rules and operator
product expansions, where it parameterizes nonperturbative corrections
to short-distance correlation functions
\cite{Shifman:1978bx,Shifman:1978by}. Its existence demonstrates that
the QCD vacuum contains coherent gluonic fields even in the absence of
external hadrons.

%In the instanton picture developed in Sec.~\ref{sec:vacuum}, these
%fluctuations arise from the same topological vacuum structure
%responsible for the axial anomaly. The scalar gluonic operator entering
%the trace anomaly and the pseudoscalar topological density entering the
%axial anomaly therefore probe complementary aspects of the same
%nonperturbative gauge dynamics

The trace anomaly also enters directly into the thermodynamics of
strongly interacting matter. At finite temperature,
\begin{equation}
\epsilon - 3P
=
\langle T^\mu_{\ \mu}\rangle_T ,
\label{eq:interaction_measure}
\end{equation}
measures deviations from conformal behavior and provides a sensitive
probe of deconfinement and chiral restoration
\cite{Nowak:1996aj,Boyd:1996bx,Borsanyi:2010cj}.
Lattice calculations show that the interaction measure peaks near the
confinement transition, indicating strong nonperturbative dynamics even
above the critical temperature~\cite{Nowak:1996aj,Hansson:2021slq}.

%=====================================================

\subsection{Instanton contribution to the gluon condensate}

The trace anomaly admits a particularly transparent semiclassical
interpretation in the instanton liquid model discussed in
Sec.~\ref{sec:vacuum}. Since the action of a single instanton is fixed
by topology,
\begin{equation}
S_{\rm inst}
=
\frac14
\int d^4x\,
G_{\mu\nu}^aG_{\mu\nu}^a
=
\frac{8\pi^2}{g^2},
\end{equation}
each pseudoparticle contributes a fixed amount to the scalar gluon
operator. For an ensemble of instantons and anti-instantons with
density \(n\), 
\begin{equation}
\left\langle
G_{\mu\nu}^aG_{\mu\nu}^a
\right\rangle_E
\simeq
\frac{32\pi^2}{g^2}\,n ,
\label{eq:G2_instanton_density}
\end{equation}
up to corrections from instanton interactions and finite packing
fraction effects
\cite{Shuryak:1981ff,Diakonov:1995ea,Schafer:1996wv,Nowak:1996aj}.

Substituting this result into the trace anomaly,
and using the one-loop beta function
gives the estimate
\begin{equation}
\left\langle
T^\mu_{\ \mu}
\right\rangle
\simeq
-b\,n ,
\label{eq:trace_instanton}
\end{equation}
with 
\[b=
\frac{11}{3}N_c-\frac{2}{3}N_f .\]
The negative sign reflects the asymptotic freedom of QCD. Using
Lorentz invariance,
\begin{equation}
\langle T_{\mu\nu}\rangle
=
\epsilon_{\rm vac}\,g_{\mu\nu},
\qquad
\langle T^\mu_{\ \mu}\rangle
=
4\epsilon_{\rm vac},
\end{equation}
one obtains
\begin{equation}
\epsilon_{\rm vac}
\simeq
-\frac{b}{4},n .
\label{eq:vacuum_energy_instanton}
\end{equation}

The instanton liquid therefore provides a simple semiclassical picture
of the trace anomaly: the same topological gauge configurations that
generate chirality-changing processes and contribute to the axial
anomaly also induce a nonzero gluon condensate and a negative vacuum
energy density associated with the breaking of scale invariance.

%====================================================

\subsection{Heat-kernel and spectral interpretation}

To see the emergence of the trace anomaly from the effective action at one-loop, 
we can use the  background-field method, and write
\[
A_\mu=\bar A_\mu+a_\mu ,
\]
The integration over the quantum fluctuations \(a_\mu\) and the
Faddeev-Popov ghosts, yields schematically
\begin{widetext}
\begin{equation}
\Gamma^{(1)}[\bar A]
=
\frac12 \log\det \Delta_{\mu\nu}^{\rm gluon}
-
\log\det \Delta^{\rm ghost}
-
N_f\log\det(i\slashed D) .
\end{equation}
\end{widetext}
The first two terms receive contributions from the gauge-fixed Yang-Mills theory, while the last term from the fermions.
The ultraviolet divergences of these determinants are obtained from
their heat-kernel expansions,
\[
{\rm Tr}\,e^{-s\Delta}
\sim
\frac{1}{(4\pi s)^2}
\sum_{n=0}^{\infty} a_n(\Delta)\,s^n ,
\qquad s\rightarrow0 .
\]
In four dimensions the logarithmic divergence is controlled by
\(a_2\). For the combined gluon, ghost, and quark operators this gives
\[
\Gamma_{\rm div}^{(1)}
=
\frac{1}{4}
\frac{1}{16\pi^2}
\left(
\frac{11}{3}N_c-\frac{2}{3}N_f
\right)
\ln\Lambda^2
\int d^4x\,
\bar G^a_{\mu\nu}\bar G^{a\mu\nu}.
\]
Equivalently, this divergence renormalizes the gauge coupling and
produces
\[
\beta(g)
=
-\frac{g^3}{16\pi^2}
\left(
\frac{11}{3}N_c-\frac{2}{3}N_f
\right)
+\cdots .
\]
The trace anomaly then follows from the Weyl and RG Ward identities.
Thus the heat-kernel perspective is spectral, but not specifically a
Dirac-spectral effect. The trace anomaly is governed by the
short-proper-time coefficients of all covariant fluctuation operators:
gluons, ghosts, and quarks.

This should be contrasted with the axial anomaly. There the anomaly is
controlled by the index of the Dirac operator and the spectral
asymmetry of its zero modes. The trace anomaly instead reflects the
logarithmic ultraviolet divergence of the full spectrum. Thus both
anomalies admit heat-kernel representations, but they probe different
parts of the spectrum: the axial anomaly is governed by topology and
zero modes, whereas the trace anomaly is governed by the ultraviolet
density of states and the resulting renormalization-group flow
\cite{Fujikawa:1980rc,DeWitt:1965jb,Seeley:1967ea,Gilkey:1975iq,
Vassilevich:2003xt}.

%====================================================

%====================================================

%====================================================

\subsection{Effective and holographic perspectives}

At long distances, the breaking of scale invariance can be incorporated
into effective descriptions through a scalar mode associated with
fluctuations of the gluon condensate. This scalar degree of freedom is
often referred to as a dilaton or conformal compensator field.

Introducing a scalar field $\chi$ with scaling dimension one, one may
construct an effective potential constrained by the anomaly,
\begin{equation}
V(\chi)
\sim
\chi^4
\left(
\ln\frac{\chi}{\Lambda_{\rm QCD}}
-
\frac14
\right).
\label{eq:dilaton_potential}
\end{equation}
The stationary condition
\begin{equation}
\frac{dV}{d\chi}=0
\label{eq:dilaton_stationary}
\end{equation}
generates a nonzero vacuum expectation value,
\begin{equation}
\langle\chi\rangle
\sim
\Lambda_{\rm QCD},
\label{eq:dilaton_vev}
\end{equation}
thereby realizing dimensional transmutation at the effective level.
Effective dilaton models have been widely applied in studies of hadron
structure, dense matter, and the interplay between chiral and conformal
symmetry breaking
\cite{Migdal:1982jp,Crewther:2013vea,Golterman:2016lsd,Brown:1991kk}.

A complementary interpretation arises in holographic approaches to QCD.
In gauge/gravity duality, the scalar gluonic operator entering the trace
anomaly is dual to a bulk scalar field propagating in a
five-dimensional curved background. Exact conformal symmetry
corresponds to pure Anti-de Sitter geometry, while the running QCD
coupling induces departures from AdS through the radial evolution of
the bulk dilaton field.

In this framework, the beta function is geometrized: the holographic
radial coordinate corresponds to the renormalization scale, and the
trace anomaly emerges from the near-boundary behavior of the metric and
scalar fields. The expectation value of the trace of the
energy-momentum tensor is then obtained through holographic
renormalization, providing a geometric realization of dimensional
transmutation and confinement
\cite{Maldacena:1997re,Witten:1998qj,Gubser:1998bc,
Erlich:2005qh,Karch:2006pv}.

%========================================
\section{Mass identities and the nucleon trace anomaly}
\label{sec:massidentities}
%========================================

The trace anomaly connects directly to hadron masses through matrix
elements of the energy-momentum tensor. For a one-nucleon state
$|P\rangle$ normalized as
\begin{equation}
\langle P|P'\rangle
=
2E_P(2\pi)^3\delta^{(3)}(P-P'),
\label{eq:nucleon_normalization}
\end{equation}
Lorentz invariance  implies
\begin{equation}
\langle P|
T^{\mu\nu}
|P\rangle
=
2P^\mu P^\nu .
\label{eq:EMT_matrix_element}
\end{equation}
Taking the trace gives
\begin{equation}
\langle P|
T^\mu_{\ \mu}
|P\rangle
=
2M_N^2 .
\label{eq:nucleon_trace_identity}
\end{equation}

This relation follows solely from Poincaré symmetry and the
normalization of one-particle states. Since the trace vanishes
classically in the chiral limit, the nonzero nucleon mass immediately
signals the importance of quantum effects and the breaking of scale
invariance through the trace anomaly.

Using the trace anomaly relation already derived in
Eq.~(\ref{eq:full_trace_anomaly}), one obtains
\begin{widetext}
\begin{equation}
2M_N^2
=
\left\langle P\left|
\frac{\beta(g)}{2g}
G_{\mu\nu}^aG^{a\mu\nu}
+
\sum_{f=1}^{N_f}
m_f(1+\gamma_m)\bar\psi_f\psi_f
\right|P\right\rangle .
\label{eq:bulk_mass_identity}
\end{equation}
\end{widetext}
This identity makes explicit that the nucleon mass originates from two
qualitatively different sources: explicit quark masses and dynamically
generated gluonic contributions associated with confinement and the
trace anomaly. Numerically, the quark-mass contribution accounts for
only a small fraction of the proton mass, with the dominant component
arising from nonperturbative gluonic fields and vacuum dynamics
\cite{Ji:1995sv,Yang:2018nqn,Ji:2021mtz,Roberts:2021nhw}.

The scalar gluonic operator entering the anomaly probes fluctuations of
the vacuum gluon condensate, while the scalar quark operator measures
the response of the nucleon to explicit chiral symmetry breaking.

Within the instanton vacuum, the gluonic matrix element probes
fluctuations of the instanton ensemble itself. In semiclassical
approaches, quark zero modes propagating through the instanton
background generate both spontaneous chiral symmetry breaking and a
substantial fraction of the low-energy nucleon mass
\cite{Shuryak:1981ff,Diakonov:1995ea,
Schafer:1996wv,Diakonov:2002fq,Nowak:1996aj,Zahed:2021fxk}.

The trace anomaly relation also has important implications for modern
studies of hadron structure. Since the energy-momentum tensor couples
directly to gravity, its matrix elements define gravitational form
factors that encode the spatial distribution of mass, pressure, and
shear forces inside hadrons
\cite{Polyakov:2018zvc,Polyakov:2019lbq,Hackett:2023rif,Ji:1994av,Nowak:1996aj}.

%====================================================

\begin{figure}[H]
\centering
\includegraphics[width=\textwidth,height=0.5\textheight,keepaspectratio]{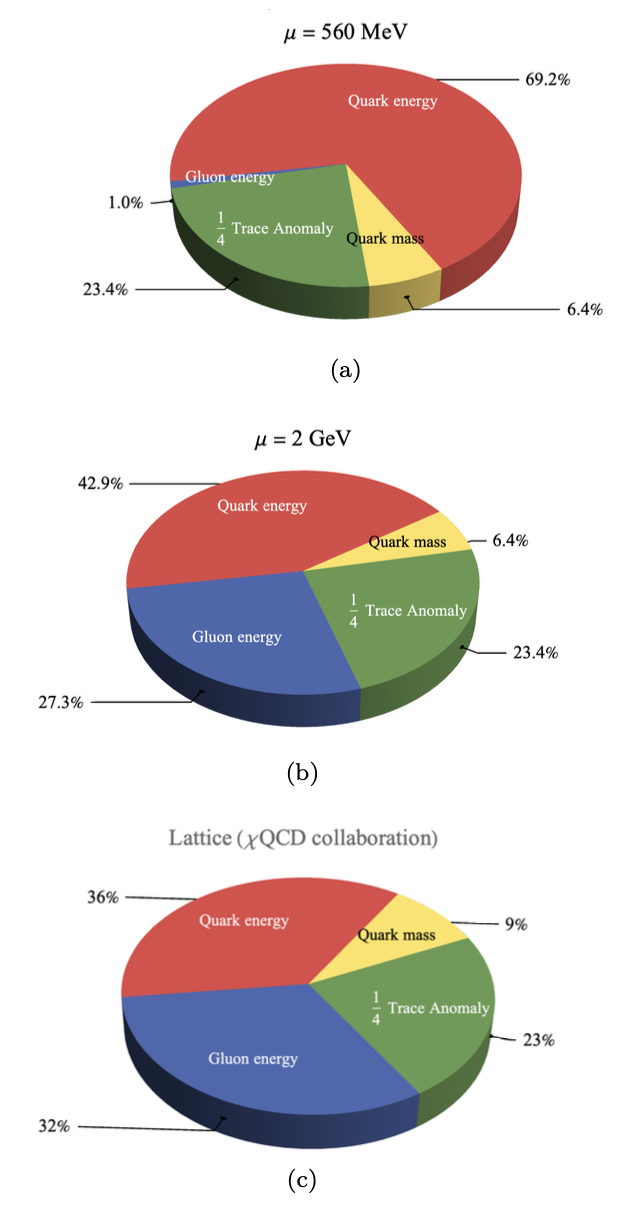}
\caption{
Nucleon mass decomposition based on Ji's gauge-invariant mass sum rule~\cite{Ji:1994av,Ji:1995sv}.
Panel (a) shows the decomposition in the QCD instanton vacuum at the
low normalization scale $\mu \simeq 560~{\rm MeV}$, where the dominant
contribution arises from quark zero modes propagating through the
topological vacuum background. Panel (b) shows the same decomposition
after DGLAP evolution to the higher scale
$\mu \simeq 2~{\rm GeV}$, where gluonic contributions increase through
perturbative QCD evolution. Panel (c) displays lattice-QCD results at
the same scale for comparison. For more details see~\cite{Liu:2024rdm}.}
\label{fig:mass_decomposition}
\end{figure}

Fig.~\ref{fig:mass_decomposition} illustrates an important conceptual aspect of hadron mass:
mass decompositions are resolution dependent. At low scales, the
dominant degrees of freedom are strongly dressed quarks propagating
through the nonperturbative vacuum, whereas at higher scales
perturbative gluons and sea quarks become increasingly important.

%====================================================

\subsection{Ji decomposition and resolution dependence}

The trace identity constrains the total hadron mass but does not specify
how that mass is distributed among quark, gluon, and anomalous
contributions. A more detailed decomposition follows from Ji's
gauge-invariant separation of the energy-momentum tensor into traceless
and trace components~\cite{Ji:1994av,Ji:1995sv},
\begin{equation}
T^{\mu\nu}
=
\bar T^{\mu\nu}
+
\hat T^{\mu\nu},
\qquad
\hat T^{\mu\nu}
=
\frac14
g^{\mu\nu}
T^\alpha_{\ \alpha}.
\label{eq:Ji_EMT_split}
\end{equation}
The nucleon mass may then be decomposed schematically as
\begin{equation}
M_N
=
M_N^Q
+
M_N^G
+
M_N^A
+
M_N^m ,
\label{eq:Ji_mass_decomposition}
\end{equation}
where the terms correspond to quark kinetic energy, gluon field energy,
anomalous trace contributions, and explicit quark masses
\cite{Ji:1994av,Ji:1995sv}.

The traceless part of the energy-momentum tensor contains the quark
and gluon kinetic and potential energy densities, while the trace part
contains the anomaly and explicit mass breaking terms. Since the trace anomaly
is proportional to the QCD beta function, the anomalous contribution
vanishes in a conformal theory but remains substantial in QCD because
of dimensional transmutation.

In the semiclassical instanton vacuum, the dominant low-resolution
contribution originates from quark zero modes propagating through the
topological medium, while direct gluonic contributions are suppressed
by the small packing fraction of the instanton ensemble. After
perturbative DGLAP evolution to higher resolution scales, momentum and
mass are redistributed toward gluonic degrees of freedom
\cite{Qian:2015wyq,Zahed:2021fxk}.

This scale dependence illustrates an important conceptual point:
decompositions of hadron mass depend on the renormalization scale and
factorization scheme. The total mass is fixed by the trace anomaly,
while the partition into quark and gluon components evolves with
resolution. Similar considerations arise in nucleon spin
decompositions, where quark and gluon angular momentum contributions
also mix under renormalization
\cite{Ji:1996ek,Leader:2013jra}.

Current  lattice-QCD calculations increasingly provide quantitative
determinations of the various mass components through matrix elements
of the energy-momentum tensor and gravitational form factors
\cite{Yang:2018nqn,Shanahan:2018nnv,Alexandrou:2020fca}.

%%%%%%%%%%%%%%%%%%%%%%%%%%%%%%%%%%%%%%%%%%%%%%%%%%%%%%%%%%%%%%%%%%%%%%%%%%%%%%
\subsection{Recent developments in mass decomposition.}
%%%%%%%%%%%%%%%%%%%%%%%%%%%%%%%%%%%%%%%%%%%%%%%%%%%%%%%%%%%%%%%%%%%%%%%%%%%%%%

The precise identification of the trace-anomaly contribution to the
hadron mass has received considerable recent attention. Beyond the
original decomposition in terms of quark, gluon, quark-mass, and trace
anomaly contributions, several complementary formulations have
reexamined the separation of the energy-momentum tensor into traceless
and trace components, the role of operator mixing under
renormalization, and the interpretation of gravitational form factors
\cite{Hatta:2018sqd,Metz:2020vxd,Lorce:2021niq}. Closely related
analyses have also established analogous mass sum rules in QED,
highlighting the universal role of the trace anomaly in relativistic
quantum field theory \cite{Rodini:2020gqh}. These developments are
consistent with the general picture presented here, while providing more discussions on the renormalization and operator structure underlying the hadron mass decomposition.

%====================================================

\subsection{Vacuum compressibility and gluonic response}

In the instanton-liquid picture, fluctuations of the scalar gluonic
operator
\(
G^2\equiv G^a_{\mu\nu}G^{a\mu\nu}
\)
measure fluctuations of the local instanton density. Through the
gluonic part of the trace anomaly,
\begin{equation}
\theta_G
\equiv
T^\mu_{\ \mu}\big|_G
=
\frac{\beta(g)}{2g}G^2
\simeq
-\frac{b}{32\pi^2}G^2 ,
\label{eq:thetaG_trace_anomaly}
\end{equation}
with  to one-loop
\[
b
=
\frac{11}{3}N_c
-
\frac{2}{3}N_f .
\]
These fluctuations are equivalently fluctuations of the gluonic vacuum
energy density. In terms of the  connected correlator
\begin{equation}
\Pi_{\theta_G}(x)
=
\left\langle
\theta_G(x)\theta_G(0)
\right\rangle_C .
\label{eq:thetaG_correlator}
\end{equation}
the corresponding gluonic compressibility may be written as
\begin{equation}
\sigma_{\theta_G}
=
\frac{1}{\langle \theta_G\rangle}
\int d^4x\,
\Pi_{\theta_G}(x).
\label{eq:thetaG_compressibility}
\end{equation}
Using Eq.~(\ref{eq:thetaG_trace_anomaly}), this becomes
\begin{equation}
\sigma_{\theta_G}
=
-\frac{b}{32\pi^2}
\int d^4x\,
\frac{
\left\langle
G^2(x)G^2(0)
\right\rangle_C
}{
\langle G^2\rangle
}.
\label{eq:thetaG_compressibility_G2}
\end{equation}
or equivalently
\begin{equation}
\sigma_{G^2}
=
\frac{1}{32\pi^2}
\int d^4x\,
\frac{
\left\langle
G^2(x)G^2(0)
\right\rangle_C
}{
\langle G^2\rangle
},
\label{eq:sigmaG2_def}
\end{equation}
with
\begin{equation}
\sigma_{\theta_G}
=
-b\,\sigma_{G^2}.
\label{eq:sigma_thetaG_sigmaG2}
\end{equation}
In the dilute instanton liquid,
\begin{equation}
\sigma_{G^2}
\simeq
\frac{4}{b},
\qquad
\sigma_{\theta_G}
\simeq
-4 .
\label{eq:sigmaG2_thetaG_result}
\end{equation}

This relation links vacuum compressibility directly to
renormalization-group dynamics. The response of the vacuum to scalar
perturbations is therefore governed by the same beta function that
controls asymptotic freedom and the trace anomaly.

The scalar gluonic correlator is also related to low-energy theorems
for the energy-momentum tensor and to scalar glueball-like
excitations in the vacuum
\cite{Novikov:1981xj,Shifman:1988zk,Liu:2024rdm}. Because the trace anomaly couples
directly to scalar gluonic fields, fluctuations of the vacuum energy
density naturally induce scalar collective modes associated with the
breaking of scale invariance.

Measurements of gluonic matrix elements in hadrons therefore provide
information not only about hadron structure but also about the
collective response of the nonperturbative QCD vacuum itself
\cite{Polyakov:2018zvc,Polyakov:2019lbq,Burkert:2018bqq,Zahed:2021fxk,Liu:2024rdm}.
%====================================================

%========================================
\section{Nucleon Spin and Anomalous Currents in QCD}
\label{sec:spin}
%========================================

One of the most important developments in current hadron physics is the
realization that the spin of the nucleon cannot be understood simply as
the sum of the spins of three valence quarks. This is in contrast with the nucleon magnetic moment which can be described as the sum of the constituent magnetic moments paving the way to the original constituent quark model. In the naive constituent
quark model, the proton spin is expected to be carried predominantly by
quark helicities,
\begin{equation}
\Delta \Sigma \simeq 1 .
\label{eq:naive_spin_sum}
\end{equation}

Polarized deep inelastic scattering experiments, however, revealed that
the quark helicity contribution is substantially smaller,
\begin{equation}
\Delta \Sigma \sim 0.2-0.3 ,
\label{eq:measured_DeltaSigma}
\end{equation}
showing that a large fraction of the nucleon spin must arise from gluon
helicity and orbital angular momentum
\cite{Ashman:1987hv,Deur:2018roz}.

This discovery, commonly referred to as the proton spin puzzle,
highlighted the fundamentally relativistic and gauge-theoretic nature
of spin in QCD. As discussed previously, the singlet axial current
mixes with gluonic operators through the axial anomaly, implying that
quark helicity is not separately conserved
\cite{Kodaira:1979pa,Altarelli:1988nr,Jaffe:1989jz}.

In this section we examine the operator structure of angular momentum
in QCD and the relation between the various spin decompositions used in
modern hadronic physics, emphasizing the role of gauge invariance,
renormalization, anomalous currents
\cite{Ji:1996ek,Leader:2013jra,Ji:2020ena}, and the role of the QCD vacuum~\cite{Zahed:2022wae}.

%====================================================

\subsection{Angular momentum in gauge theory}

In relativistic field theory, angular momentum follows from Noether's
theorem associated with Lorentz invariance. Starting from the
energy-momentum tensor $T^{\mu\nu}$, the conserved angular momentum
current is
\begin{equation}
M^{\mu\nu\rho}
=
x^\nu T^{\mu\rho}
-
x^\rho T^{\mu\nu},
\label{eq:angular_momentum_current}
\end{equation}
with conserved generators
\begin{equation}
J^{\nu\rho}
=
\int d^3x\,
M^{0\nu\rho}(x).
\label{eq:angular_momentum_generator}
\end{equation}
For a nucleon state $|P,S\rangle$ normalized covariantly,
\begin{equation}
\langle P,S|P',S'\rangle\nonumber
=
2E_P(2\pi)^3
\delta^{(3)}(P-P')\delta_{SS'},
\label{eq:spin_state_normalization}
\end{equation}
rotational invariance implies
\begin{equation}
\langle P,S|J^i|P,S\rangle
=
\frac12\,S^i .
\label{eq:spin_expectation}
\end{equation}

The central issue in QCD is not the conservation of total angular
momentum, but its decomposition into quark and gluon contributions.
Because gluons carry both spin and orbital motion, and because canonical
momentum operators are generally not gauge invariant, the decomposition
of nucleon spin is not unique
\cite{Jaffe:1989jz,Ji:1996ek}.

Different operator definitions correspond to different ways of
separating spin and orbital motion between matter and gauge fields.
These distinctions become especially important in non-Abelian gauge
theories because gluons themselves carry angular momentum and interact
nonlinearly through the gauge structure.

%====================================================

\subsection{Spin decomposition in QCD}

At the formal level, the nucleon spin sum rule may be written as
\begin{equation}
\frac12
=
J_q
+
J_g ,
\label{eq:total_spin_sum}
\end{equation}
where $J_q$ and $J_g$ denote the total angular momentum carried by
quarks and gluons.
The quark contribution is decomposed into helicity and orbital parts,
\begin{equation}
J_q
=
\frac12\Delta\Sigma
+
L_q ,
\label{eq:quark_spin_decomposition}
\end{equation}
while the gluon contribution is written schematically as
\begin{equation}
J_g
=
\Delta G
+
L_g .
\label{eq:gluon_spin_decomposition}
\end{equation}
Here $\Delta\Sigma$ denotes the  intrinsic spin contribution or net quark helicity  summed
over flavors,
\begin{equation}
\Delta\Sigma
=
\sum_{f=1}^{N_f}\Delta q_f ,
\label{eq:DeltaSigma_sum}
\end{equation}
$\Delta G$ the gluon helicity, and $L_q$, $L_g$ the corresponding
orbital angular momentum contributions.

These quantities are not uniquely defined observables. Their operator
definitions depend on the treatment of gauge fields, Wilson lines, and
the distinction between canonical and kinetic momentum operators
\cite{Leader:2013jra,Lorce:2017wkb}.

Nucleon spin is also intrinsically scale dependent. Renormalization
continuously redistributes angular momentum between quark and gluon
sectors as the resolution scale changes
\cite{Ji:1995sv,Qian:2015wyq,Zahed:2022wae,Liu:2024rdm}.

%====================================================

\subsection{Singlet axial current and anomalous helicity flow}

The quark helicity contribution is related to the singlet axial current
through the nucleon matrix element.
As discussed earlier, the singlet axial current satisfies the anomalous
divergence relation given in Eq.~(\ref{eq:axialanomalyintro}). The appearance
of the topological gluonic operator implies that the singlet axial
charge mixes with gluonic degrees of freedom under renormalization.
Consequently, $\Delta\Sigma$ depends on the renormalization scale and
factorization scheme
\cite{Kodaira:1979pa,Altarelli:1988nr}.

This mixing appears explicitly in polarized deep inelastic scattering.
For the first moment of the polarized structure function,
% Eq.~(\ref{eq:g1_moments}),
%\begin{equation}
%\Gamma_1(Q^2)
%=
%\int_0^1dx\, g_1(x,Q^2),
%\label{eq:Gamma1_spin}%\nonumber
%\end{equation}
the operator product expansion gives Eq.(\ref{eq:g1_moments}).
The anomaly therefore allows axial charge to flow between quark and
gluon sectors. This transfer is ultimately tied to the topology of
gauge fields and the nonperturbative vacuum structure of QCD
\cite{Schafer:1996wv,Nowak:1996aj,Diakonov:2002fq,Zahed:2022wae}.

The EMC measurements demonstrated experimentally that the quark
helicity contribution is much smaller than expected from constituent
quark models, motivating extensive studies of gluon polarization and
orbital angular momentum
\cite{Ashman:1987hv,Deur:2018roz}.

%====================================================

\paragraph{Ji decomposition}

A particularly important decomposition was introduced by Ji using the
symmetric gauge-invariant Belinfante energy-momentum tensor~\cite{Ji:1996ek}. In this
approach the nucleon spin is split as in Eq.~(\ref{eq:total_spin_sum})
where both contributions are manifestly gauge invariant.
The quark and gluon angular momenta are defined through nucleon matrix
elements of the QCD energy-momentum tensor,
\begin{align}
J_q
&=
\frac12
\left[
A_q(0)+B_q(0)
\right],
\label{eq:Jq_gravitational}
\\
J_g
&=
\frac12
\left[
A_g(0)+B_g(0)
\right].
\label{eq:Jg_gravitational}
\end{align}
Equivalently, Ji's relation may be expressed through generalized
parton distributions~\cite{Ji:1996ek},
\begin{equation}
J_q
=
\frac12
\int_{-1}^{1}dx\,
x\,
\left[
H_q(x,\xi,0)
+
E_q(x,\xi,0)
\right].
\label{eq:Ji_GPD_relation}
\end{equation}

A major strength of Ji's decomposition is that it is gauge invariant
and directly connected to measurable quantities through deeply virtual
Compton scattering and lattice QCD calculations
\cite{Ji:1996ek,Ji:1998pc,Diehl:2003ny}. However, the gluon
contribution corresponds to the total gluon angular momentum and does
not separate gluon spin and gluon orbital motion in a local,
manifestly gauge-invariant way.

The same matrix elements also determine gravitational form factors,
which encode the spatial distributions of mass, pressure, and shear
forces inside hadrons
\cite{Polyakov:2018zvc,Polyakov:2019lbq}. Modern lattice-QCD
calculations increasingly determine these quantities directly from QCD
\cite{Yang:2018nqn,Alexandrou:2020fca}.

%====================================================

\paragraph{Jaffe-Manohar decomposition}

A different decomposition was proposed by Jaffe and Manohar using the
canonical angular momentum tensor derived directly from the QCD
Lagrangian. In this framework the nucleon spin is decomposed as~\cite{Jaffe:1989jz}
\begin{equation}
\frac12
=
\frac12\Delta\Sigma
+
\Delta G
+
L_q^{\rm can}
+
L_g^{\rm can}.
\label{eq:Jaffe_Manohar_sum}
\end{equation}
The advantage of this decomposition is that it separates quark and
gluon spin contributions explicitly and admits a natural partonic
interpretation in light-front quantization. In light-cone gauge,
\(
A^+ = 0 ,
\label{eq:lightcone_gauge}
\)
the gluon helicity operator reduces schematically to
\begin{equation}
\Delta G
\sim
\int d^3x\,
\left(
\mathbf E^a
\times
\mathbf A^a
\right)^3 .
\label{eq:gluon_helicity_operator}
\end{equation}

This decomposition is therefore especially useful in analyses of
high-energy scattering processes and polarized parton distributions
\cite{Jaffe:1989jz,Brodsky:1997de}. However, the canonical operators
are not manifestly gauge invariant. Beyond light-cone gauge,
Wilson-line structures must be introduced, and the separation between
spin and orbital motion becomes path dependent.

The light-front formulation underlying this decomposition provides a
direct connection to parton distribution functions and Fock-space wave
functions.

%====================================================

\paragraph{Relation between Ji and Jaffe-Manohar}

The difference between the Ji and Jaffe-Manohar decompositions can be
traced to the distinction between kinetic and canonical momentum in a
gauge theory.
In the Ji decomposition, orbital motion is constructed from the
gauge-covariant derivative,
\begin{equation}
D_\mu
=
\partial_\mu
-
igA_\mu ,
\label{eq:covariant_derivative_spin}\nonumber
\end{equation}
leading to kinetic orbital angular momentum operators. By contrast, the
Jaffe-Manohar decomposition uses canonical derivatives involving only
$\partial_\mu$.
Schematically,
\begin{equation}
\mathbf p_{\rm kin}
=
-i\mathbf D
=
-i\nabla
-
g\mathbf A ,
\label{eq:kinetic_momentum}
\end{equation}
while the canonical momentum contains only the derivative term.
Consequently, the two orbital angular momentum operators differ by a
gauge-potential contribution often referred to as potential angular
momentum
\cite{Hatta:2011ku,Hatta:2012cs}.

The Ji decomposition emphasizes gauge-invariant energy-momentum flow
and is naturally suited for lattice calculations and generalized
parton-distribution analyses. The Jaffe-Manohar decomposition
emphasizes the partonic interpretation of spin in the
infinite-momentum frame and is closely connected to light-front wave
functions and polarized scattering observables.

Modern gauge-invariant extensions separate the gauge field into
pure-gauge and physical components,
\(
A_\mu
=
A_\mu^{\rm pure}
+
A_\mu^{\rm phys},
%\label{eq:Chen_decomposition}
\)
allowing canonical-like operators to be rewritten in a gauge-invariant
form
\cite{Chen:2008ag,Wakamatsu:2010cb}.
The two decompositions therefore represent different but complementary
organizations of the same underlying QCD angular momentum structure.

%====================================================

%====================================================

\subsection{Orbital motion and phase-space structure}

The decomposition of orbital angular momentum is closely related to the
phase-space structure of the nucleon. In modern formulations, quark
and gluon orbital motion can be expressed through generalized
transverse-momentum distributions and Wigner distributions
\cite{Lorce:2011kd,Meissner:2009ww,Hatta:2011ku}.

Schematically, quark orbital angular momentum may be written as
\begin{equation}
L_q
\sim
\int
dx\,
d^2k_\perp\,
d^2b_\perp\,
(\mathbf b_\perp\times\mathbf k_\perp)_z
\,W_q(x,\mathbf b_\perp,\mathbf k_\perp),
\label{eq:OAM_Wigner_distribution}
\end{equation}
where $W_q$ denotes the quark Wigner distribution.
These phase-space distributions interpolate between ordinary parton
distributions, transverse-momentum distributions, and form factors,
providing a multidimensional description of nucleon structure. They
also clarify how orbital motion emerges dynamically from confinement
and gauge interactions.

The appearance of Wigner distributions provides a direct connection
between nucleon spin physics and quantum phase-space methods familiar
from many-body theory. In QCD, however, gauge invariance requires the
inclusion of Wilson lines, implying that orbital motion is inherently
linked to the gauge structure of the vacuum and to color transport
through the hadron~\cite{Burkardt:2012sd,Hatta:2012cs}.

Recent experimental programs at Jefferson Lab, COMPASS, RHIC, and the
future Electron-Ion Collider aim to constrain these multidimensional
distributions through deeply virtual Compton scattering, exclusive
meson production, and polarized semi-inclusive scattering
\cite{Accardi:2012qut,AbdulKhalek:2021gbh}.

\subsection{Vacuum topology and the flavor-singlet axial charge}

\subsubsection{Vacuum topology and spin structure}

The axial anomaly implies that nucleon spin cannot be understood purely
in terms of constituent quark degrees of freedom. Because the singlet
axial current mixes with the topological gluonic operator appearing in
Eq.~(\ref{eq:axialanomalyintro}), vacuum topology contributes directly to the
helicity structure of the nucleon.

In semiclassical approaches based on the instanton liquid, chirality
fluctuations induced by instantons continuously transfer axial charge
between quarks and gluons
\cite{Schafer:1996wv,Nowak:1996aj,Diakonov:2002fq,Zahed:2022wae}. Quark zero modes
propagating through the topological vacuum reduce the net quark
helicity carried by valence degrees of freedom and generate nontrivial
orbital motion.

From this perspective, the proton spin puzzle reflects not simply
missing constituent spin contributions but the fundamentally collective
and topological nature of angular momentum in QCD. The nucleon spin
emerges from the interplay of quark helicity, gluon polarization,
orbital motion, and vacuum topology. The anomaly thereby links
measurable spin observables directly to the global topological
structure of gauge fields and to the spectral properties of the Dirac
operator discussed earlier through the index theorem
\cite{Atiyah:1968mp,Nielsen:1981hk}.

Modern analyses of polarized scattering increasingly suggest that
gluon helicity and orbital angular momentum contributions become
important at short distances or high resolution, while nonperturbative vacuum effects
dominate at low resolution
\cite{Qian:2015wyq,Zahed:2022wae,Liu:2024rdm,Shuryak:2026pqt}. The evolution between
these regimes is controlled by perturbative QCD evolution equations and
anomaly-induced operator mixing
\cite{Kodaira:1979pa,Altarelli:1988nr}.

The topological viewpoint also connects nucleon spin physics to broader
phenomena involving anomalous transport and chirality flow in QCD. In
hot and dense matter, axial-charge fluctuations induced by topology can
generate observable effects such as the chiral magnetic effect and
related anomalous transport phenomena
\cite{Kharzeev:2015znc,Kharzeev:2007jp,Fukushima:2008xe}.

More generally, topology-induced chirality fluctuations provide a
direct bridge between hadron structure and the dynamics of strongly
interacting matter under extreme conditions. The same anomalous
currents appearing in polarized scattering reappear in the
hydrodynamic description of quark matter, indicating that
anomaly-induced transport is a universal feature of non-Abelian gauge
theories
\cite{Son:2009tf,Landsteiner:2016led}.

%====================================================

\subsubsection{Axial current, anomaly, and intrinsic quark spin}

The intrinsic quark spin contribution to the nucleon is directly tied
to the singlet axial current discussed previously in
Sec.~\ref{sec:axialanom}. The gauge-invariant singlet current defined
earlier  satisfies the anomalous
Ward identity already given in Eq.~(\ref{eq:axialanomalyintro}).

For a polarized nucleon state $|P,S\rangle$, the singlet axial current
defines the intrinsic quark spin contribution through
\begin{equation}
\langle P',S|
J_5^\mu
|P,S\rangle
=
2M_N\,\Sigma_Q^N(q^2)\,
S^\mu ,
\label{eq:SigmaQ_definition}
\end{equation}
where
\begin{equation}
\Sigma_Q^N(0)\equiv\Delta\Sigma .
\label{eq:SigmaQ_forward}
\end{equation}
Taking the divergence and using the nucleon Dirac equation gives
\begin{equation}
q_\mu
\langle P',S|
J_5^\mu
|P,S\rangle
=
2M_N\,\Sigma_Q^N(q^2)\,
\bar N_S(P')
\,i\gamma_5\,
N_S(P).
\label{eq:divergence_axial_matrix}
\end{equation}

The anomalous Ward identity then implies
\begin{widetext}
\begin{equation}
2M_N\,\Sigma_Q^N(q^2)\,
\bar N_S(P')
i\gamma_5
N_S(P)
=
\left[
\frac{g^2N_f}{16\pi^2}
A_G(q^2)
+
2\sum_{f=1}^{N_f}
m_f A_P^{(f)}(q^2)
\right]
\bar N_S(P')
i\gamma_5
N_S(P),
\label{eq:Ward_identity_formfactors}
\end{equation}
\end{widetext}
where the form factors are defined as
\begin{align}
\langle P',S|
G_{\mu\nu}^a\tilde G^{a\mu\nu}
|P,S\rangle
&=
A_G(q^2)\,
\bar N_S(P')
i\gamma_5
N_S(P),
\label{eq:AG_definition}
\\
\langle P',S|
\bar\psi_f i\gamma_5\psi_f
|P,S\rangle
&=
A_P^{(f)}(q^2)\,
\bar N_S(P')
i\gamma_5
N_S(P).
\label{eq:AP_definition}
\end{align}
Taking the forward limit yields
\begin{equation}
\Delta\Sigma
=
\frac{g^2N_f}{32\pi^2M_N}
A_G(0)
+
\frac{1}{M_N}
\sum_{f=1}^{N_f}
m_f A_P^{(f)}(0).
\label{eq:DeltaSigma_forward_relation}
\end{equation}
In the chiral limit, the pseudoscalar contribution becomes negligible
and the intrinsic quark spin is controlled directly by the topological
gluonic operator,
\begin{equation}
\Delta\Sigma
\simeq
\frac{g^2N_f}{32\pi^2M_N}
A_G(0).
\label{eq:DeltaSigma_chiral}
\end{equation}
This relation provides a direct bridge between the spin carried by
quarks and the topological structure of the QCD vacuum.

%====================================================

\subsubsection{Topological fluctuations in the instanton vacuum}

In the instanton liquid model developed earlier, the axial charge
fluctuates through changes in the topological charge,
\(
Q
=
N_+
-
N_- ,
\)
where $N_+$ and $N_-$ denote the numbers of instantons and
anti-instantons, respectively.
The relevant matrix element of the topological density is obtained from
the connected three-point function
\begin{widetext}
\begin{equation}
\frac{
\langle P,S|
G_{\mu\nu}^a\tilde G^{a\mu\nu}(0)
|P,S\rangle
}{
\langle P,S|P,S\rangle
}
=
\lim_{T\to\infty}
\frac{
\left\langle
J_P^\dagger(T)\,
G_{\mu\nu}^a\tilde G^{a\mu\nu}(0)\,
J_P(-T)
\right\rangle_C
}{
\left\langle
J_P^\dagger(T)
J_P(-T)
\right\rangle
},
\label{eq:GdualG_threepoint}
\end{equation}
\end{widetext}
 where the subscript $C$ denotes the connected contribution. Here $J_P(-T)$ is a nucleon source in the Euclidean past $-T$, ans $J\dagger (T)$ a nucleon sink in the
 Euclidean future $T$.

In a canonical ensemble the total topological charge is fixed and the
connected contribution vanishes identically. A nonzero result arises
only in the grand-canonical ensemble, where fluctuations of $Q$ are
allowed. One then finds schematically~\cite{Diakonov:1995qy,Kacir:1996qn,Zahed:2022wae}
\begin{widetext}
\begin{equation}
\frac{V}{32\pi^2}
\frac{
\langle P,S|
G_{\mu\nu}^a\tilde G^{a\mu\nu}
|P,S\rangle
}{
\langle P,S|P,S\rangle
}
\simeq
\langle Q^2\rangle
\frac{\partial}{\partial Q}
\log
\left[
\lim_{T\to\infty}
\left\langle
J_P^\dagger(T)
J_P(-T)
\right\rangle
\right],
\label{eq:GdualG_Q_derivative}
\end{equation}
\end{widetext}
Using the asymptotic behavior of the nucleon correlator,
\begin{equation}
\left\langle
J_P^\dagger(T)
J_P(-T)
\right\rangle
\sim
e^{-2M_NT},
\label{eq:nucleon_correlator_asymptotic}
\end{equation}
one obtains
\begin{equation}
\frac{V}{32\pi^2}
\frac{
\langle P,S|
G_{\mu\nu}^a\tilde G^{a\mu\nu}
|P,S\rangle
}{
M_N
\langle P,S|P,S\rangle
}
\simeq
-
\chi_{V}
\left(
\frac{\partial\log M_N(Q)}
{\partial Q}
\right)_{Q=0},
\label{eq:GdualG_mass_response}
\end{equation}
where
\(
\chi_V=\langle Q_V^2\rangle/V
\)
denotes the topological charge fluctuation density in a finite
four-volume $V$. with 
the infinite volume limit given in
Eq.~(\ref{eq:chi_top_theta}).
Equation~(\ref{eq:GdualG_mass_response}) shows explicitly that the
intrinsic quark spin is tied to the response of the nucleon mass to
fluctuations of vacuum topology.

%====================================================

\subsubsection{Intrinsic quark spin}

To estimate the effect quantitatively, one may use a quark-diquark
picture of the nucleon,
\begin{equation}
p_\uparrow
\sim
u_\uparrow[ud]_0 ,
\label{eq:quark_diquark_proton}
\end{equation}
in which the spin is carried primarily by the unpaired u-quark, as the dominant diquark configuration in the nucleon is a spin-scalar.
In the instanton vacuum, quark propagation occurs through hopping
between instanton zero modes. The constituent quark mass generated by
the instanton ensemble therefore depends on the topological background.
For a dilute instanton liquid model,
\begin{equation}
M_N(Q)
\simeq
M_N
-
M_u(0)\,
s^\uparrow
\frac{Q}{\bar N},
\label{eq:MNQ_expansion}
\end{equation}
where $M_u(0)$ is the dynamical constituent quark mass, hence
\begin{equation}
\left(
\frac{\partial\log M_N(Q)}
{\partial Q}
\right)_{Q=0}
\simeq
-
\frac{M_u(0)}{M_N}
\frac{s^\uparrow}{\bar N}.
\label{eq:dlogMN_dQ}
\end{equation}

%+++++++++
Substituting this relation into
Eq.~(\ref{eq:GdualG_mass_response}) yields
\begin{equation}
\Delta\Sigma
\sim
\left(
\frac{\chi_V}{n}
\right)
\left(
\frac{M_u(0)}{M_N}
\right),
\label{eq:DeltaSigma_ILM}
\end{equation}
 In the instanton vacuum, the small-volume limit is
Poissonian
\cite{Shuryak:1994rr}
\begin{equation}
\lim_{V\rightarrow0}
\frac{\langle Q_V^2\rangle}{V}
\simeq n,
\label{eq:small_volume_limit}
\end{equation}
showing that local topological fluctuations are controlled by the
instanton plus anti-instanton density, with the estimate~\cite{Zahed:2022wae}
\(
\Delta\Sigma
\simeq 0.6,
\)
consistent with the observed suppression of the flavor-singlet axial
charge relative to the naive constituent-quark expectation
\cite{Zahed:2022wae,Liu:2024rdm}. The result suggests that the relevant
topological fluctuations governing the propagation of singlet axial
charge are local fluctuations within finite spacetime domains rather
than the screened thermodynamic susceptibility.

%====================================================

\subsection{Gluon helicity and spin phenomenology}

A simple qualitative insight follows from the instanton description of
the QCD vacuum. The gluon angular momentum appearing in gauge-invariant
spin decompositions is associated with the Poynting-vector structure
\begin{equation}
\vec J_g
\sim
\int d^3x\,
\vec x
\times
(
\vec E^a
\times
\vec B^a
).
\label{eq:Jg_Poynting}
\end{equation}
For a self-dual instanton field,
\begin{equation}
\vec E^a
=
\pm
\vec B^a ,
\end{equation}
so that
\begin{equation}
\vec E^a
\times
\vec B^a
=
0 .
\end{equation}
Consequently, an isolated instanton does not contribute directly to the
gluon angular momentum. Nonvanishing contributions arise only through
multi-instanton interactions and are therefore suppressed by the small
instanton packing fraction
\begin{equation}
\kappa
=
\pi^2 \bar\rho^4 n
\simeq
0.1 .
\end{equation}

\begin{figure}[H]\centering\includegraphics[width=\textwidth,height=0.5\textheight,keepaspectratio]{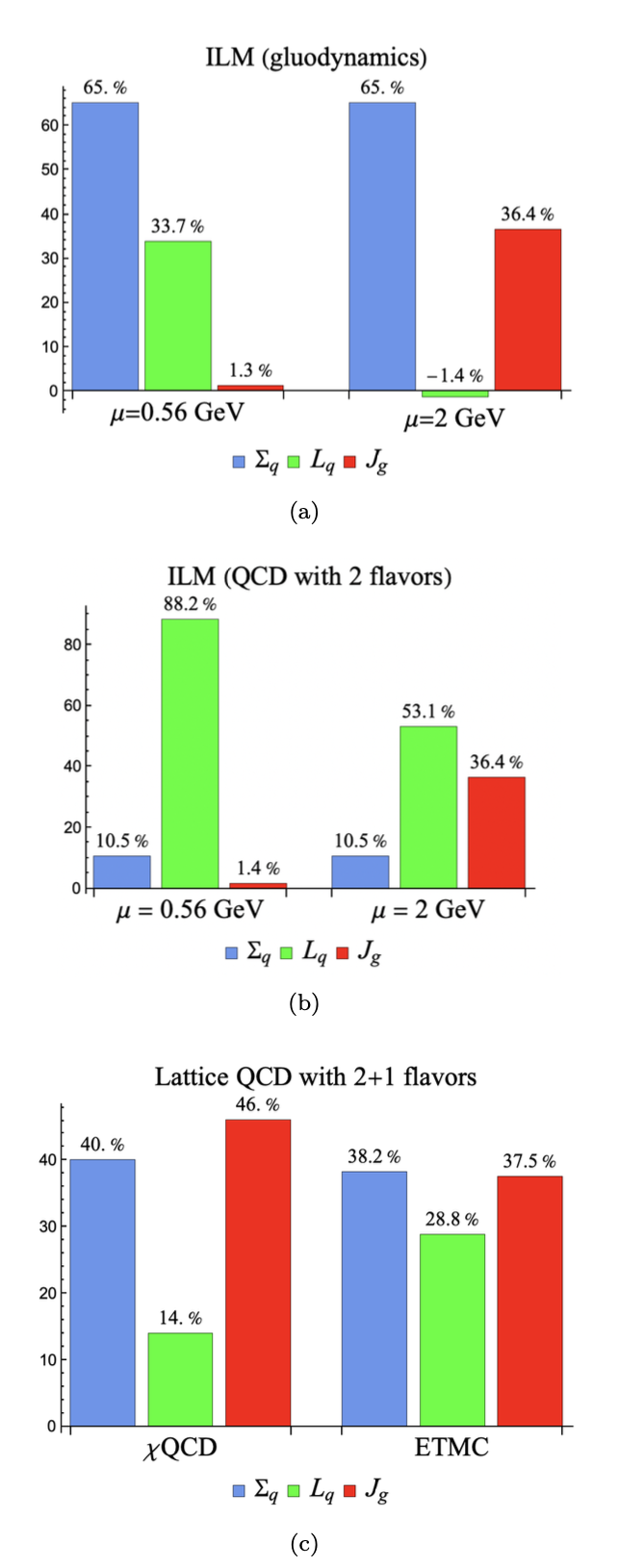}\caption{Nucleon spin decomposition in the instanton vacuum and comparison withlattice QCD from~\cite{Liu:2024rdm}. Panel (a) shows the decomposition in pure gluodynamics,where the spin structure is dominated by topological gauge fields.Panel (b) displays the full QCD instanton vacuum including dynamicalquarks and orbital motion generated through quark zero modes. Panel (c)shows lattice-QCD results at comparable normalization scales.}
\label{fig:spin}
\end{figure}

This picture suggests that at the low normalization scale associated
with the instanton vacuum, the nucleon spin is dominated by quark spin
and orbital motion, while the gluon contribution remains relatively
small. A representative decomposition obtained in the instanton vacuum
is shown in Fig.~\ref{fig:spin} and compared with recent
lattice-QCD results \cite{Liu:2024rdm}. Under perturbative evolution to
higher momentum scales, gluon helicity and gluon orbital angular
momentum increase through operator mixing and radiative corrections
\cite{Leader:2013jra,Deur:2018roz}.

%====================================================

%====================================================

%========================================
\section{High-Energy QCD, the Operator Product Expansion, and Anomalies}
\label{sec:OPE}
%========================================

The previous sections showed how anomalies, topology, and vacuum
structure govern many nonperturbative aspects of QCD, including chiral
symmetry breaking, hadron masses, and nucleon spin structure. These
same phenomena also appear in high-energy processes, where
short-distance perturbative dynamics can be systematically separated
from long-distance hadronic structure.

Deep inelastic scattering (DIS) provides the primary experimental probe
of QCD at short distances. In this regime a highly virtual photon
resolves quark and gluon constituents inside the nucleon at distance
scales much smaller than the confinement scale. The theoretical
framework underlying DIS combines perturbative QCD, renormalization
group evolution, and the operator product expansion (OPE), allowing
observable structure functions to be related to matrix elements of
local operators
\cite{Wilson:1969zs,Gross:1973id,Politzer:1973fx}.

The coefficient functions appearing in the OPE are determined by
short-distance perturbative dynamics, while the operator matrix
elements encode long-distance confinement physics, vacuum condensates,
and topological structure. In polarized scattering, the axial anomaly
provides the connection between perturbative evolution and
nonperturbative gluonic topology
\cite{Kodaira:1979pa,Altarelli:1988nr,Jaffe:1989jz}.

The experimental access to these spin-dependent matrix elements is provided by polarized deep inelastic scattering, to which we now turn

%====================================================

\subsection{Deep inelastic scattering}

The inclusive DIS process is
\begin{equation}
\ell(k)+N(P)
\rightarrow
\ell(k')+X ,
\label{eq:DIS_process}
\end{equation}
where $X$ denotes the inclusive hadronic final state. The interaction
proceeds through exchange of a virtual photon with momentum
\(
q^\mu
=
k^\mu-k'^\mu ,
\label{eq:virtual_photon_momentum}
\)
and virtuality
\(
Q^2
=
-q^2 >0 .
\label{eq:DIS_virtuality}
\)
The Bjorken scaling variable is
\begin{equation}
x_B
=
\frac{Q^2}{2P\cdot q},
\label{eq:Bjorken_x}
\end{equation}
which in the parton model corresponds to the longitudinal momentum
fraction carried by the struck quark.
The deep inelastic limit is defined by
\begin{equation}
Q^2\rightarrow\infty,
\qquad
\nu=\frac{P\cdot q}{M_N}\rightarrow\infty,
\qquad
x_B \ {\rm fixed}.
\label{eq:Bjorken_limit}
\end{equation}
In this regime asymptotic freedom implies that the virtual photon
interacts over short distances
\(x^2\sim \frac{1}{Q^2},\)
allowing perturbative methods to be applied to the hard scattering
subprocess.

Within the parton model, the virtual photon scatters incoherently from
quarks carrying momentum fraction $x_B$ inside the fast-moving nucleon
\cite{Feynman:1969ej}. Bjorken scaling corresponds to the approximate
independence of structure functions on $Q^2$, while logarithmic scaling
violations arise from perturbative gluon radiation and renormalization
group evolution in QCD.

The separation between short- and long-distance physics forms the basis
of factorization in perturbative QCD
\cite{Collins:1989gx,Ellis:1991qj}.

%====================================================

\subsection{Hadronic tensor and structure functions}

The central object in DIS is the hadronic tensor,
\begin{equation}
W_{\mu\nu}
=
\frac{1}{4\pi}
\int d^4x\,
e^{iq\cdot x}
\langle P,S|
[J_\mu(x),J_\nu(0)]
|P,S\rangle ,
\label{eq:hadronic_tensor}
\end{equation}
where the electromagnetic current is
\begin{equation}
J_\mu
=
\sum_{f=1}^{N_f}
e_f
\bar\psi_f\gamma_\mu\psi_f .
\label{eq:electromagnetic_current_DIS}
\end{equation}
Lorentz covariance, parity, and current conservation allow the tensor
to be decomposed into symmetric and antisymmetric parts,
\begin{equation}
W_{\mu\nu}
=
W_{\mu\nu}^{(S)}
+
W_{\mu\nu}^{(A)} .
\label{eq:hadronic_tensor_decomposition}
\end{equation}

For polarized scattering the antisymmetric component is
\begin{widetext}
\begin{equation}
W_{\mu\nu}^{(A)}
=
\frac{i}{P\cdot q}
\epsilon_{\mu\nu\alpha\beta}
q^\alpha
\left[
S^\beta g_1(x_B,Q^2)
+
\left(
S^\beta
-
\frac{S\cdot q}{P\cdot q}P^\beta
\right)
g_2(x_B,Q^2)
\right],
\label{eq:polarized_hadronic_tensor}
\end{equation}
\end{widetext}
where $g_1$ and $g_2$ are polarized structure functions.
The structure function $g_1$ probes helicity distributions of quarks
and gluons inside the nucleon, while $g_2$ contains higher-twist
information associated with quark-gluon correlations and transverse
spin dynamics.

The symmetric part defines the unpolarized structure functions $F_1$
and $F_2$, which in the naive parton model satisfy the Callan-Gross
relation
\begin{equation}
F_2(x_B,Q^2)
=
2x_BF_1(x_B,Q^2),
\label{eq:Callan_Gross}
\end{equation}
reflecting the spin-$\frac12$ nature of quark constituents
\cite{Callan:1969uq}.
Moments of the structure functions are related through the OPE to
matrix elements of local operators. In particular, the first moment of
$g_1$ probes axial-current matrix elements and is therefore sensitive
to anomaly-induced operator mixing and gluonic topology
\cite{Kodaira:1979pa,Jaffe:1989jz}. The structure function $g_2$ is related to the Lorentz force~\cite{Aslan:2019jis,Liu:2025ypg}.

%====================================================

\subsection{Operator product expansion}

The operator product expansion provides the framework for analyzing the
short-distance behavior of the current product appearing in the
hadronic tensor. Near the light cone,
\(
x^2\rightarrow0,
\label{eq:lightcone_limit}
\)
the time-ordered product of currents admits an expansion in local
operators,
\begin{equation}
T\,
J_\mu(x)J_\nu(0)
\sim
\sum_n
C_{\mu\nu}^{(n)}(x,\mu)
\,O_n(0,\mu).
\label{eq:OPE_current_product}
\end{equation}
The Wilson coefficients encode short-distance perturbative physics and
may be computed systematically in powers of $\alpha_s(Q^2)$, while the
local operators contain the long-distance nonperturbative structure of
the nucleon.

The separation between short- and long-distance dynamics is controlled
by factorization,
\begin{widetext}
\begin{equation}
[{\rm observable}]
=
[{\rm coefficient\ function}]
\otimes
[{\rm matrix\ element}],
\label{eq:factorization_formula}
\end{equation}
\end{widetext}
which is one of the central organizing principles of perturbative QCD.
Operators are classified by their twist,
\(
\tau
=
d-s ,
\label{eq:twist_definition}
\)
where $d$ is the canonical dimension and $s$ the Lorentz spin.
Leading-twist operators dominate the Bjorken limit, while higher-twist
operators are suppressed by powers of $1/Q^2$.

For polarized structure functions, the relevant leading-twist operators
are axial-vector operators,
\begin{equation}
O_{5}^{\mu_1\cdots\mu_n}
=
\sum_{f=1}^{N_f}
\bar\psi_f
\gamma^{\{\mu_1}
\gamma_5
iD^{\mu_2}\cdots iD^{\mu_n\}}
\psi_f ,
\label{eq:twist2_axial_operator}
\end{equation}
where braces denote symmetrization and subtraction of traces.
Their matrix elements determine moments of polarized parton
distributions and evolve logarithmically with the renormalization scale
through anomalous dimensions
\cite{Gross:1974cs}.

The renormalization-group evolution of Wilson coefficients and
operators leads directly to the DGLAP evolution equations governing
parton distributions
\cite{Dokshitzer:1977sg,Gribov:1972ri,Altarelli:1977zs}. In polarized
scattering, the singlet axial current mixes with gluonic operators
through the anomaly, producing scale-dependent helicity transfer
between quark and gluon sectors.

%====================================================

\subsection{Moments and local operators}

A major consequence of the OPE is the relation between moments of
structure functions and matrix elements of local operators. Mellin
moments convert the nonlocal light-cone expansion into a tower of local
operators with definite spin and twist
\cite{Wilson:1969zs,Gross:1974cs}.

For polarized scattering,
\begin{equation}
\int_0^1dx\,
x^{n-1}
g_1(x,Q^2)
=
\sum_i
C_n^{(i)}(Q^2/\mu^2,\alpha_s)
\,
a_n^{(i)}(\mu),
\label{eq:g1_moments}
\end{equation}
where
\begin{equation}
a_n^{(i)}
=
\langle P,S|
O_n^{(i)}
|P,S\rangle .
\label{eq:reduced_matrix_elements}
\end{equation}
The scale dependence of both Wilson coefficients and operator matrix
elements is constrained by the renormalization group so that physical
observables remain independent of the factorization scale $\mu$.

The forward matrix elements of the twist-two axial operators introduced
in Eq.~(\ref{eq:twist2_axial_operator}) are parameterized as
\begin{equation}
\langle P,S|
O_{5\,\mu_1\cdots\mu_n}^{(f)}
|P,S\rangle
=
2a_n^{(f)}
S_{\{\mu_1}
P_{\mu_2}\cdots P_{\mu_n\}},
\label{eq:operator_matrix_element_param}
\end{equation}
which defines the reduced matrix elements entering polarized parton
moments.
For the lowest moment,
\(
n=1,
\label{eq:first_moment}
\)
the operator reduces to the axial current discussed previously in
Sec.~\ref{sec:spin}. Polarized DIS therefore directly probes the
singlet and nonsinglet axial-current matrix elements governing nucleon
spin structure.

In the naive parton model, the reduced matrix elements are related to
polarized quark distributions through
\begin{equation}
a_n^{(f)}
=
\int_0^1dx\,
x^{n-1}
\left[
\Delta q_f(x)
+
(-1)^{n-1}
\Delta\bar q_f(x)
\right].
\label{eq:an_parton_moments}
\end{equation}
Thus the tower of twist-two operators encodes the complete set of
moments of polarized parton distributions. Modern lattice-QCD
calculations increasingly determine these matrix elements directly from
first principles
\cite{Alexandrou:2020fca,Constantinou:2020hdm}.

%====================================================

\subsection{Perturbative evolution and anomalous dimensions}

Although Bjorken scaling emerges approximately in the parton model,
QCD predicts logarithmic scaling violations due to radiative
corrections. These effects are governed by renormalization-group
evolution and provide one of the central quantitative tests of
asymptotic freedom
\cite{Gross:1973id,Politzer:1973fx}.

The Wilson coefficients and local operators satisfy
\begin{equation}
\mu
\frac{d}{d\mu}
O_n
=
-\gamma_n
O_n ,
\label{eq:operator_RG_evolution}
\end{equation}
where $\gamma_n$ are anomalous dimensions. Equivalently, the Wilson
coefficients evolve according to
\begin{equation}
\left[
\mu\frac{\partial}{\partial\mu}
+
\beta(g)\frac{\partial}{\partial g}
-
\gamma_n
\right]
C_n
=
0 .
\label{eq:Wilson_coefficient_RG}
\end{equation}

The anomalous dimensions arise from ultraviolet divergences in loop
corrections and determine how operators evolve with resolution scale.
In Mellin space they are related directly to moments of the splitting
functions appearing in the DGLAP equations.
For polarized parton distributions, the scale dependence is governed by
\begin{align}
\frac{\partial}{\partial\ln Q^2}
\Delta q_f(x,Q^2)
&=
\sum_{f'=1}^{N_f}
\Delta P_{qq}^{ff'}
\otimes
\Delta q_{f'}
+
\Delta P_{qg}
\otimes
\Delta G ,
\label{eq:DGLAP_quark_polarized}
\\
\frac{\partial}{\partial\ln Q^2}
\Delta G(x,Q^2)
&=
\sum_{f=1}^{N_f}
\Delta P_{gq}^{f}
\otimes
\Delta q_f
+
\Delta P_{gg}
\otimes
\Delta G .
\label{eq:DGLAP_gluon_polarized}
\end{align}
The convolution symbol denotes
\begin{equation}
(P\otimes f)(x)
=
\int_x^1\frac{dy}{y}\,
P\!\left(\frac{x}{y}\right)
f(y),
\label{eq:convolution_definition}
\end{equation}
reflecting collinear parton splitting through perturbative radiation.

These equations demonstrate explicitly that quark and gluon helicities
mix under evolution. The mixing is perturbative in origin but is tied
physically to the axial anomaly and anomalous operator mixing.
At leading order, the polarized splitting functions describe helicity
transfer through quark and gluon emission processes. The kernels
receive substantial higher-order radiative corrections and become
factorization-scheme dependent
\cite{Altarelli:1977zs,Dokshitzer:1977sg,Mertig:1995ny,
Vogelsang:1995vh}.
Perturbative evolution therefore describes how nucleon spin is
redistributed among quark and gluon degrees of freedom as the
resolution scale changes.

%====================================================

\subsection{Axial anomaly in the OPE}

The axial anomaly modifies the operator product expansion in the singlet
channel because the divergence of the singlet axial current contains the
topological operator introduced previously.
As discussed earlier, the singlet axial current satisfies the anomalous
Ward identity given in Eq.~(\ref{eq:axialanomalyintro}). Consequently, the
singlet axial current mixes with gluonic operators under
renormalization. In polarized DIS this leads schematically to
\begin{equation}
a_0(Q^2)
=
\Delta\Sigma(Q^2)
-
\frac{N_f\alpha_s(Q^2)}{2\pi}
\Delta G(Q^2),
\label{eq:singlet_axial_charge_OPE}
\end{equation}
where $a_0(Q^2)$ is the singlet axial charge extracted experimentally.

The decomposition of the singlet axial current depends on the
factorization scheme. In the Adler-Bardeen scheme the anomaly is
contained entirely in the coefficient function, whereas in the
$\overline{\rm MS}$ scheme part of the anomalous contribution is
absorbed into the singlet quark distribution
\cite{Adler:1969gk,Bardeen:1969md,Altarelli:1988nr,
Kodaira:1998jn}.

Equation~(\ref{eq:singlet_axial_charge_OPE}) shows explicitly that the
anomaly provides a mechanism through which perturbative evolution
becomes sensitive to nonperturbative gluonic structure.

The anomaly also plays an important role in the small-$x$ behavior of
polarized structure functions, where gluon helicity effects may become
increasingly significant. Modern approaches combining perturbative
resummation, saturation physics, and helicity evolution aim to
describe this high-energy regime systematically
\cite{Kovchegov:2015pbl,Kovchegov:2016weo}.

%====================================================

\subsection{Bjorken and Ellis-Jaffe sum rules}

An important application of the OPE is the derivation of sum rules for
polarized structure functions. These relations connect experimentally
measured moments of structure functions directly to symmetry generators
and conserved currents in QCD.

The Bjorken sum rule relates the difference of proton and neutron
polarized structure functions to the nonsinglet axial charge,
\begin{widetext}
\begin{equation}
\int_0^1dx\,
\left[
g_1^p(x,Q^2)
-
g_1^n(x,Q^2)
\right]
=
\frac{1}{6}
g_A
\left[
1
-
\frac{\alpha_s(Q^2)}{\pi}
+\cdots
\right].
\label{eq:Bjorken_sum_rule}
\end{equation}
\end{widetext}
Because the nonsinglet axial current is anomaly free, the Bjorken sum
rule provides a particularly clean test of perturbative QCD and has
been experimentally verified with high precision
\cite{Bjorken:1966jh,Deur:2004ti}.
Perturbative corrections to the Bjorken sum rule have been computed to
high orders in QCD
\cite{Larin:1991tj,Baikov:2010je}.

By contrast, the Ellis-Jaffe sum rule involves the singlet axial
current and is therefore sensitive to anomaly effects and gluonic
topology. Naively one expects
\begin{equation}
a_0
\simeq
\Delta\Sigma ,
\label{eq:Ellis_Jaffe_naive}
\end{equation}
leading to the expectation that quark helicities carry most of the
proton spin.
Experimentally, however,
\(
\Delta\Sigma
\simeq
0.3 ,
\label{eq:DeltaSigma_DIS}
\)
much smaller than expected from simple quark models
\cite{Ashman:1987hv,Filippone:2001ux} as we noted earlier.

%This discrepancy led to the proton spin puzzle and demonstrated the
%importance of anomalous helicity transfer between quark and gluon
%sectors.

%====================================================

%\subsection{Interplay between perturbative and nonperturbative QCD}

%%%%%%%%%%%Although DIS probes short-distance physics, the operator matrix
%%%%%%%%%%elements entering the OPE remain fundamentally nonperturbative.
%%%%%%%%%Parton distributions, generalized parton distributions, and helicity
%%%%%%%%distributions encode confinement dynamics and vacuum structure. In the
%%%%%%%singlet channel, the axial anomaly ties these quantities directly to
%%%%%%gluonic topology through the topological density operator introduced
%%%%%earlier in Eq.~(\ref{eq:topological_density}).

%%%%From the instanton perspective discussed previously, fluctuations of
%%%topological charge continuously transfer axial charge between quarks
%%and gluons. The same topological susceptibility governing the $\eta'$
%mass therefore also influences polarized parton distributions and the
%scale dependence of the singlet axial charge
%\cite{Shuryak:1981ff,Diakonov:1985eg}.

%%The OPE therefore provides a synthesis of perturbative and
%nonperturbative physics. Wilson coefficients are governed by
%short-distance asymptotic freedom, while operator matrix elements probe
%vacuum condensates, confinement, and topology
%\cite{Shifman:1978bx,Shifman:1978by}.

%%%%%%%%%%%%%%%%%%%%%%%%%%%%%%%%%%%%%%%%%%%%%
%%%%%%%%%%

\subsection{Triangle Anomaly, Anomaly Poles, and Worldline Interpretation}
\label{subsec:triangle_poles}

The anomalous contribution to polarized deep inelastic scattering is
most directly encoded in the flavor-singlet axial-vector-vector (AVV)
correlator
\cite{Adler:1969gk,Bell:1969ts,Rosenberg:1962pp}
\begin{widetext}
\begin{equation}
T_{\mu\alpha\beta}(k_1,k_2)
=
i^2
\int d^4x\,d^4y\,
e^{ik_1\cdot x+k_2\cdot y}
\langle0|
T
J_{\mu5}^{(0)}(0)
J_\alpha(x)
J_\beta(y)
|0\rangle ,
\label{eq:AVV_correlator}
\end{equation}
\end{widetext}
where \(J_{\mu5}^{(0)}\) is the flavor-singlet axial current and
\(J_\alpha\) the electromagnetic current. Gauge invariance requires
\begin{equation}
k_1^\alpha T_{\mu\alpha\beta}
= k_2^\beta T_{\mu\alpha\beta}
=
0 ,
\label{eq:AVV_vector_WI}
\end{equation}
while the divergence of the axial current is fixed by the Adler-Bell-Jackiw anomaly.

The AVV amplitude admits the Rosenberg decomposition
\cite{Rosenberg:1962pp},
\begin{widetext}
\begin{align}
T_{\mu\alpha\beta}
=
F_1
\,\epsilon_{\mu\alpha\beta\rho}k_1^\rho
+
F_2
\,\epsilon_{\mu\alpha\beta\rho}k_2^\rho
+
F_3
\,k_{1\beta}
\epsilon_{\mu\alpha\rho\sigma}
k_1^\rho k_2^\sigma
+
F_4
\,k_{2\beta}
\epsilon_{\mu\alpha\rho\sigma}
k_1^\rho k_2^\sigma
+
F_5
\,k_{1\alpha}
\epsilon_{\mu\beta\rho\sigma}
k_1^\rho k_2^\sigma
+
F_6
\,k_{2\alpha}
\epsilon_{\mu\beta\rho\sigma}
k_1^\rho k_2^\sigma .
\label{eq:Rosenberg}
\end{align}
\end{widetext}
The vector Ward identities reduce the number of independent form
factors and leave a longitudinal component fixed exactly by the
anomaly,
\begin{equation}
T^{(L)}_{\mu\alpha\beta}
=
\frac{q_\mu}{q^2}
{\cal A}_{\alpha\beta}(k_1,k_2),
\qquad
q=k_1+k_2 .
\label{eq:longitudinalAVV}
\end{equation}
The factor \(1/q^2\) signals a massless anomaly pole
\cite{Dolgov:1971ri,Crewther:1972kn,
Giannotti:2008cv,Coriano:2012pe}.

The anomaly pole represents long-range propagation of axial charge.
Its residue is completely fixed by the anomaly coefficient and is
therefore protected by the Adler-Bardeen theorem
\cite{Adler:1969er,Bardeen:1969md}. In the nonsinglet channel the pole
is saturated by the pion in the chiral limit through PCAC. In the
flavor-singlet channel, however, the anomaly couples the axial current
to topological gluonic degrees of freedom, leading to the screening
mechanism responsible for the large \(\eta'\) mass
\cite{tHooft:1976rip,Witten:1979vv,Veneziano:1979ec}.

A particularly useful interpretation is provided by the worldline
formalism. The AVV correlator is represented as a path integral over
closed fermionic trajectories carrying spin degrees of freedom
\cite{Strassler:1992zr,Schubert:2001he}. The insertion of
\(\gamma_5\) selects the periodic Grassmann sector of the worldline
path integral, whose fermionic zero modes generate the
Levi-Civita tensor and reproduce the anomaly, as we showed earlier In this language the
anomaly pole reflects the long-range propagation associated with these
topological zero-mode configurations and provides a geometric
realization of the same spectral-flow mechanism that underlies the
Atiyah-Singer index theorem.

This interpretation has recently been exploited by Tarasov and
Venugopalan in a worldline analysis of polarized deep inelastic
scattering \cite{Tarasov:2021yll,Tarasov:2025mvn}. There the anomaly
pole embedded in the AVV triangle controls the leading contribution to
the polarized structure function \(g_1\), linking the proton spin
problem directly to topological screening and the infrared dynamics of
the flavor-singlet axial channel.

%++++++++++++++++++++++++++++++++++++++++++++++++++

%====================================================

\subsection{Topological Screening, Vacuum Topology, and the Flavor-Singlet Axial Charge}
\label{subsec:topological_screening}

The anomaly pole discussed above determines the coupling of the
flavor-singlet axial current to topological gauge fields. The next
question is how the corresponding axial charge propagates through the
QCD vacuum. This issue is central to the interpretation of polarized
deep inelastic scattering and to the longstanding proton spin puzzle.

The relevant object is the flavor-singlet axial-current correlator
\begin{equation}
\Pi_{\mu\nu}^{00}(q)
=
i\int d^4x\,e^{iq\cdot x}
\langle0|
T J_{\mu5}^{(0)}(x)
J_{\nu5}^{(0)}(0)
|0\rangle .
\label{eq:Pi00}
\end{equation}
Lorentz covariance allows the decomposition
\begin{equation}
\Pi_{\mu\nu}^{00}(q)
=
-g_{\mu\nu}\Pi_1(q^2)
+
q_\mu q_\nu \Pi_2(q^2),
\label{eq:Pi00decomp}
\end{equation}
where the longitudinal amplitude is sensitive to anomalous chiral Ward
identities and to the propagation of flavor-singlet axial charge.

The flavor-singlet channel differs qualitatively from the nonsinglet
sector because the axial current couples directly to topological gauge
configurations. The corresponding dynamics are encoded in the
correlation function of the topological charge density
\begin{equation}
\chi(q^2)
=
i\int d^4x\,
e^{iq\cdot x}
\langle0|
T\,q(x)\,q(0)
|0\rangle ,
\label{eq:topsusc}
\end{equation}
whose spectral representation may be written as
\begin{equation}
\chi(q^2)
=
\frac{1}{\pi}
\int_0^\infty ds,
\frac{{\rm Im}\chi(s)}
{s-q^2-i\epsilon}.
\label{eq:topspectral}
\end{equation}

The same topological susceptibility that controls the $\eta'$ mass
through the Witten-Veneziano mechanism also governs the propagation
of flavor-singlet axial charge
\cite{Witten:1979vv,Veneziano:1979ec,Shore:1991dv,Shore:1992eu}.

In pure Yang-Mills theory the zero-momentum limit of
Eq.~(\ref{eq:topsusc}) defines the topological susceptibility discussed
earlier in connection with the $U(1)_A$ problem. In full QCD, however,
the quantity most relevant for polarized deep inelastic scattering is
not $\chi(0)$ itself but rather its slope at the origin,
\begin{equation}
\chi'(0)
=
\left.
\frac{d\chi(q^2)}
{dq^2}
\right|_{q^2=0}.
\label{eq:chiprimeTV}
\end{equation}
The appearance of $\chi'(0)$ follows from anomalous chiral Ward
identities relating the longitudinal part of
Eq.~(\ref{eq:Pi00decomp}) to the topological charge density.
The resulting relation was first developed by Shore and Veneziano and
has recently been reformulated and extended 
in~\cite{Shore:1991dv,Shore:1992eu,Tarasov:2021yll,Tarasov:2025mvn}.

To show the underlying physics it is useful to introduce the
properly normalized flavor-singlet decay constant $F_0$ through
\begin{equation}
\langle0|
J_{\mu5}^{(0)}
|\eta'(q)\rangle
=
iF_0\,q_\mu .
\label{eq:F0def}
\end{equation}
The coupling of the singlet axial current to the nucleon may then be
parameterized by
\begin{equation}
\langle P,S|
J_{\mu5}^{(0)}
|P,S\rangle
=
2M_N\,g_A^{(0)}\,S_\mu ,
\label{eq:gA0def}
\end{equation}
where $g_A^{(0)}$ denotes the gauge-invariant singlet axial charge.

Combining the anomalous Ward identities with the singlet
Goldberger-Treiman relation yields
\begin{equation}
2M_N g_A^{(0)}
=
\sqrt{\frac32}\,
F_0
\left(
g_{\eta' NN}
-
g_{QNN}
\right),
\label{eq:GTsinglet}
\end{equation}
where $g_{\eta' NN}$ is the $\eta'$-nucleon coupling and
$g_{QNN}$ measures the direct coupling of the topological charge
density to the nucleon.
Equation~(\ref{eq:GTsinglet}) makes explicit that the flavor-singlet
axial charge is influenced not only by quark helicity but also by
topological gluonic degrees of freedom. In the absence of topology one
would recover the familiar OZI expectation

\begin{equation}
g_A^{(0)}
\simeq
g_A^{(8)} ,
\label{eq:OZIlimit}
\end{equation}
whereas experimentally
\begin{equation}
g_A^{(0)}
\ll
g_A^{(8)} .
\label{eq:suppression}
\end{equation}

The key insight of the Shore-Veneziano mechanism is that the
suppression of the flavor-singlet axial charge arises from
topological screening of axial charge by the QCD vacuum. In the
large-$N_c$ approximation one finds schematically
\begin{equation}
\frac{g_A^{(0)}}{g_A^{(8)}}
\propto
\frac{\sqrt{\chi'(0)}}
{\sqrt{\chi'_{\rm OZI}(0)}} ,
\label{eq:SVratio}
\end{equation}
where $\chi'_{\rm OZI}(0)$ denotes the corresponding quantity in the
absence of topological screening.

The explicit form of the intrinsic spin was derived by Shore and
Veneziano and recently revisited by Tarasov and Venugopalan
\cite{Shore:1991dv,Shore:1992eu,Tarasov:2021yll,Tarasov:2025mvn},
\begin{equation}
\Delta\Sigma
=
\frac{\sqrt{2N_f}}{2M_N}\,
\sqrt{\chi'_{\rm QCD}(0)\big|_{m=0}}\,
g_{\eta_0 NN},
\label{eq:TVrelation}
\end{equation}
where $g_{\eta_0 NN}$ is the coupling of the primordial singlet
pseudoscalar field to the nucleon and
$\chi'_{\rm QCD}(0)|_{m=0}$ denotes the slope of the full-QCD
topological susceptibility in the chiral limit. Equation
(\ref{eq:TVrelation}) makes explicit that the net quark helicity
measured in polarized scattering is governed by the infrared
topological response of the QCD vacuum.

The physical origin of Eq.~(\ref{eq:TVrelation}) becomes transparent
in the worldline formulation of polarized deep inelastic scattering
developed 
in~\cite{Tarasov:2021yll,Tarasov:2025mvn}. In this approach, the leading
contribution to the polarized structure function $g_1$ is generated by
the AVV triangle anomaly embedded in the DIS box diagram. The first
moment of $g_1$ is therefore directly connected to matrix elements of
the topological charge density,
showing that polarized DIS probes the same topological dynamics that
underlie the $U(1)_A$ problem.

A chief result of this analysis is that the anomaly pole appearing
in the AVV triangle does not survive as a physical massless
excitation. In QED the AVV pole is cancelled by the corresponding PVV
triangle contribution. In QCD the cancellation is modified by
spontaneous chiral symmetry breaking and vacuum topology. The anomaly
pole is screened by a primordial flavor-singlet pseudoscalar field
whose Wess-Zumino-Witten coupling to
$\,G\widetilde G$ shifts the pole to the physical $\eta'$
mass. The resulting screening mechanism is the deep-inelastic
counterpart of the same dynamics responsible for resolving the
$U(1)_A$ problem.

Within this framework, the susceptibility $\chi(0)$ itself is not the
relevant quantity. In full QCD it vanishes in the chiral limit owing
to topological screening. Instead, the propagation of flavor-singlet
axial charge is controlled by the slope
\begin{equation}
\chi'(0)
=
\left.
\frac{d\chi(q^2)}{dq^2}
\right|_{q^2=0},
\end{equation}
which measures the infrared response of the vacuum to topological
perturbations. The suppression of $g_A^{(0)}$ relative to its OZI
expectation may therefore be understood as a consequence of
topological screening rather than a suppression of intrinsic quark
spin.

The finite-mass extension of the worldline analysis confirms the
robustness of this picture. Finite quark masses induce corrections
through the pseudoscalar channel, but the resulting modifications of
the Shore-Veneziano relation are numerically small
\cite{Tarasov:2025mvn}. The conclusion remains unchanged: polarized
deep inelastic scattering probes not only quark and gluon helicities
but also the infrared topological structure of the QCD vacuum.

Viewed from this perspective, the proton spin problem and the
$U(1)_A$ problem are different manifestations of the same anomalous
coupling between flavor-singlet axial charge and vacuum topology.
Both are controlled by the infrared response encoded in
$\chi'(0)$, linking polarized deep inelastic scattering directly to
nonperturbative topological correlations in QCD.

\subsection{Finite-volume fluctuations and a possible $x=0$ contribution}
\label{subsec:xzero_topology}

The anomalous singlet axial Ward identity implies that the
topological susceptibility vanishes in full QCD in the chiral limit,
\(\chi(0)=0 ,\)
reflecting topological charge screening. This does not eliminate
topological fluctuations altogether. The relevant quantity governing
the propagation of flavor-singlet axial charge is the slope
\(\chi'(0)\), which remains finite and nonzero
\cite{Shore:1991dv,Shore:1992eu,Tarasov:2021yll,Tarasov:2025mvn}.
Indeed,
\begin{equation}
\chi'(0)
=
-\frac{1}{8}
\int d^4x ,
x^2
\langle q(x) q(0)\rangle ,
\label{eq:chiprime_moment}
\end{equation}
showing that $\chi'(0)$ measures finite-range topological
correlations even when the integrated susceptibility vanishes.

A complementary interpretation follows from finite-volume
fluctuations. While screening suppresses the susceptibility in the
thermodynamic limit,
\begin{equation}
\chi(0)
=
\lim_{V\rightarrow\infty}
\frac{\langle Q_V^2\rangle}{V}
=
0 ,
\end{equation}
topological fluctuations remain finite inside sufficiently small
subvolumes of the vacuum~\cite{Shuryak:1994rr,Zahed:2022wae,Liu:2024rdm},
\begin{equation}
\lim_{V\rightarrow0}
\frac{\langle Q_V^2\rangle}{V}
=
n ,
\label{eq:small_volume_susc}
\end{equation}
 Thus local
topological fluctuations survive even though the global susceptibility
is screened. Since hadrons probe only a finite spacetime region, their
axial properties are sensitive to these finite-volume fluctuations,
providing a natural link between topological screening and the
suppression of the flavor-singlet axial charge.

A related issue concerns the extraction of the singlet axial charge
from polarized deep inelastic scattering. If the dispersion relation
for the spin-dependent forward Compton amplitude requires no
subtraction at infinity, the measured first moment coincides with the
gauge-invariant singlet axial charge. If a subtraction constant is
present, however,
\begin{equation}
g_A^{(0)}
=
g_A^{(0)}\Big|_{\rm pDIS}
+
{\cal C}_\infty ,
\label{eq:gA0_Cinfty}
\end{equation}
where ${\cal C}_\infty$ corresponds to a contribution localized at
Bjorken \(x=0\)
\cite{Crewther:1977ce,Efremov:1989ze,Bass:2004xa}.

The anomaly relation and the Atiyah-Singer index theorem imply that
topologically nontrivial gauge configurations induce spectral flow and
transfer axial charge between quark modes and the gauge background.
Such zero-mode contributions are naturally associated with the
\(x\rightarrow0\) region in the infinite-momentum frame. A nonzero
${\cal C}_\infty$ would therefore represent axial charge carried by
topological vacuum degrees of freedom rather than by finite-(x)
partons.

Whether ${\cal C}_\infty$ is nonzero remains an open question. If
present, it would provide a direct connection between the proton spin
problem and the topological structure of the QCD vacuum.

%%%%%%%%%%%%%%%%%%%%%%%%%%%%%%%%%%%%%%%%%%%%%%%

%========================================
\section{Summary and Outlook}
\label{sec:conclusion}
%========================================

Quantum Chromodynamics provides one of the most striking examples of
how quantum effects reshape the realization of classical symmetries.
Although the QCD Lagrangian possesses approximate chiral symmetry and,
in the massless limit, classical scale invariance, both symmetries are
profoundly modified by quantum fluctuations. The axial anomaly links
chirality to gauge-field topology, while the trace anomaly generates
the intrinsic QCD scale through dimensional transmutation. Together,
these anomalies determine essential features of the strong interaction,
including the structure of the QCD vacuum, the generation of hadron
masses, the resolution of the $U(1)_A$ problem, and the spin structure
of the nucleon.

A central theme of this review has been the role of the QCD vacuum as
an active dynamical medium rather than an empty ground state.
Topologically nontrivial gauge configurations, including instantons,
link chirality to gauge-field topology through anomalous Ward
identities. Their effects are encoded in the topological
susceptibility, the flavor-singlet axial channel, and the anomalously
large $\eta'$ mass, providing a microscopic realization of the
$U(1)_A$ problem. These same topological dynamics continue to influence
a broad range of hadronic observables.

The trace anomaly provides a complementary perspective. Quantum
breaking of classical scale invariance generates the confinement scale
$\Lambda_{\rm QCD}$ and accounts for most of the mass of hadrons and
hence visible matter. Modern studies of the energy-momentum tensor,
gravitational form factors, lattice-QCD matrix elements, and hadron
mass decompositions have revealed increasingly direct connections
between confinement dynamics, gluonic structure, and the emergence of
hadron masses. In this sense, the trace anomaly furnishes a bridge
between the microscopic gluonic degrees of freedom of QCD and the
macroscopic properties of hadronic matter.

An important message throughout this review is that anomalous dynamics
remains visible even in observables traditionally associated with
high-energy scattering. The operator-product expansion and
perturbative-QCD evolution connect short-distance processes to matrix
elements of local quark and gluon operators, while anomaly-induced
operator mixing links quark and gluon helicity contributions in
polarized deep inelastic scattering. The flavor-singlet axial current
occupies a particularly special role because it probes both the
partonic structure of the nucleon and the topological structure of the
vacuum.

This emerging connection points toward a broader hadron-parton
correspondence in QCD, in which vacuum topology, chiral symmetry
breaking, constituent-quark dynamics, and parton distributions are
viewed as different aspects of a common underlying structure
\cite{Shuryak:2026pqt}. Such a perspective may help bridge the
traditional separation between nonperturbative descriptions based on
the QCD vacuum and perturbative descriptions based on quarks and
gluons, providing a more unified picture of hadron formation and
structure.

Recent developments have further sharpened the connection between
vacuum topology and the flavor-singlet axial channel. Studies based on
anomalous Ward identities, topological screening, instanton vacuum
models, and spectral flow have emphasized the role of topological
gauge-field fluctuations in the propagation of singlet axial charge.
These investigations suggest that the proton spin problem and the
$U(1)_A$ problem are not entirely separate phenomena but may reflect
different manifestations of the same underlying topological dynamics.
At the same time, the precise relation between the full singlet axial
charge and the quantity extracted from polarized deep inelastic
scattering remains an active area of investigation, including the
possible role of contributions localized near Bjorken $x=0$.

Modern studies of hadron structure, combining generalized and
transverse-momentum dependent parton distributions, lattice QCD, and
polarized scattering experiments, are providing increasingly direct
access to the gluonic and topological structure of hadrons. Future
measurements at the Electron-Ion Collider will extend these studies
into the small-$x$ regime, where novel manifestations of anomalous
dynamics may emerge.

From a broader perspective, the axial and trace anomalies provide the
two fundamental links between the microscopic and macroscopic
descriptions of QCD. The axial anomaly ties chirality to vacuum
topology and governs the flavor-singlet sector, while the trace
anomaly connects gluonic dynamics to the emergence of mass. Together
they furnish a unified framework for understanding how hadronic
structure arises from quarks and gluons.

Many open questions remain. The quantitative relation between vacuum
topology and the flavor-singlet axial charge, the possible role of
topological contributions at Bjorken $x=0$, the interplay between
orbital angular momentum and topological gauge-field fluctuations, and
the emergence of hadron mass from gluonic dynamics continue to be
active areas of investigation.

Ultimately, anomalies provide one of the clearest windows into the
deep structure of Quantum Chromodynamics. They connect the topology of
gauge fields to the properties of hadrons, link vacuum fluctuations to
observable scattering amplitudes, and illuminate how the strong
interaction generates mass, spin, and structure from quarks and
gluons. The continuing exploration of these connections remains one of
the most fertile directions in contemporary strong-interaction
physics.

%%%%%%%%%%%%%%%%%%%%%%%%%%%%%%%%%%%%%%%%%%%%%%

{\centerline{\bf Acknowledgements}}
\vskip 0.5cm
This work is supported by the Office of Science, U.S. Department of
Energy under Contract No.~DE-FG-88ER40388.
This research is also supported in part within the framework of the
Quark-Gluon Tomography (QGT) Topical Collaboration, under contract
No.~DE-SC0023646.

%========================================
\appendix
%========================================

%========================================
\section{Technical Derivations}
\label{app:anomaly_derivations}
%========================================

This appendix collects technical derivations supporting the anomaly
relations used in the main text. Since the physical interpretation and
applications were already discussed earlier, only the essential
mathematical steps are retained here.

%====================================================

\subsection{AVV triangle anomaly}

The perturbative derivation follows the original analyses of Adler and
of Bell and Jackiw~\cite{Adler:1969gk,Bell:1969ts}.
Consider the axial-vector-vector amplitude
\begin{widetext}
\begin{equation}
T^{\mu\nu\rho}(p,q)
=
\sum_{f=1}^{N_f}
\int \frac{d^4k}{(2\pi)^4}
{\rm Tr}
\left[
\gamma^\mu\gamma_5
\frac{1}{\slashed{k}-m_f}
\gamma^\nu
\frac{1}{\slashed{k}+\slashed{p}-m_f}
\gamma^\rho
\frac{1}{\slashed{k}-\slashed{q}-m_f}
\right],
\label{eq:app_AVV_triangle}
\end{equation}
\end{widetext}
where the axial current carries momentum $p+q$.
The integral is superficially linearly divergent, so a shift of loop
momentum changes the integral by a finite surface term,
\begin{equation}
k\rightarrow k+a\,p+b\,q .
\label{eq:app_loop_shift}
\end{equation}
The vector Ward identities follow from
\begin{equation}
p_\nu\gamma^\nu
=
(\slashed{k}+\slashed{p}-m_f)-(\slashed{k}-m_f),
\label{eq:app_vector_identity}
\end{equation}
giving
\begin{widetext}
\begin{align}
p_\nu T^{\mu\nu\rho}
&=
\sum_{f=1}^{N_f}
\int\frac{d^4k}{(2\pi)^4}
{\rm Tr}
\left[
\gamma^\mu\gamma_5
\left(
\frac{1}{\slashed{k}-m_f}
-
\frac{1}{\slashed{k}+\slashed{p}-m_f}
\right)
\gamma^\rho
\frac{1}{\slashed{k}-\slashed{q}-m_f}
\right].
\label{eq:app_vector_Ward}
\end{align}
%\end{widetext}

Gauge invariance fixes the momentum routing so that
\begin{equation}
p_\nu T^{\mu\nu\rho}=0,
\qquad
q_\rho T^{\mu\nu\rho}=0 .
\label{eq:app_vector_conservation}
\end{equation}
The axial divergence then becomes
%\begin{widetext}
\begin{equation}
(p+q)_\mu\gamma^\mu\gamma_5
=
\gamma_5
\left[
(\slashed{k}+\slashed{p}-m_f)
-
(\slashed{k}-\slashed{q}-m_f)
\right]
+
2m_f\gamma_5 ,
\label{eq:app_axial_contraction}
\end{equation}
\end{widetext}
and the Dirac trace gives
\begin{equation}
{\rm Tr}
\left(
\gamma^\mu\gamma^\nu\gamma^\rho\gamma^\sigma\gamma_5
\right)
=
-4i\epsilon^{\mu\nu\rho\sigma}.
\label{eq:app_gamma5_trace}
\end{equation}

After Feynman parametrization and gauge-invariant regularization one
obtains
\begin{equation}
(p+q)_\mu T^{\mu\nu\rho}
=
\frac{N_f}{2\pi^2}
\epsilon^{\nu\rho\alpha\beta}
p_\alpha q_\beta
+
2\sum_{f=1}^{N_f}m_f\,T_{P,f}^{\nu\rho}.
\label{eq:app_AVV_divergence}
\end{equation}
Coupling to non-Abelian gauge fields yields the anomaly equation already
used in the main text,
The anomaly coefficient is exact because of the Adler-Bardeen
nonrenormalization theorem
\cite{Adler:1969gk,Bell:1969ts,Bardeen:1969md}.

%====================================================

\subsection{Fujikawa derivation from the fermion measure}

The anomaly may also be derived from the non-invariance of the fermion
measure under chiral transformations. For quarks in a background gauge
field,
\begin{equation}
Z[A]
=
\int
\prod_{f=1}^{N_f}
{\cal D}\bar\psi_f\,{\cal D}\psi_f\,
\exp\left[
i\int d^4x\,
\sum_{f=1}^{N_f}
\bar\psi_f
i\slashed D
\psi_f
\right].
\label{eq:app_partition_function}
\end{equation}
Under the infinitesimal local chiral transformation
\begin{equation}
\psi_f
\rightarrow
\psi_f+i\alpha(x)\gamma_5\psi_f,
\qquad
\bar\psi_f
\rightarrow
\bar\psi_f+i\alpha(x)\bar\psi_f\gamma_5,
\label{eq:app_chiral_rotation}
\end{equation}
the classical action changes by
\begin{equation}
\delta S
=
-\int d^4x\,
\alpha(x)\,
\partial_\mu J_5^\mu .
\label{eq:app_deltaS}
\end{equation}

Expanding the fermions in eigenfunctions of the Euclidean Dirac
operator,
\begin{widetext}
\begin{equation}
i\slashed D\,\phi_n
=
\lambda_n\phi_n,
\qquad
\psi_f(x)=\sum_n a_n^{(f)}\phi_n(x),
\qquad
\bar\psi_f(x)=\sum_n \bar b_n^{(f)}\phi_n^\dagger(x),
\label{eq:app_Dirac_eigenbasis}
\end{equation}
%\end{widetext}
the fermion measure becomes
\begin{equation}
\prod_{f=1}^{N_f}
{\cal D}\bar\psi_f\,{\cal D}\psi_f
=
\prod_{f=1}^{N_f}
\prod_n
d\bar b_n^{(f)}\,da_n^{(f)} .
\label{eq:app_measure}
\end{equation}
The chiral rotation induces the Jacobian
%\begin{widetext}
\begin{equation}
\exp\left[
-2i
\sum_{f=1}^{N_f}
\int d^4x\,
\alpha(x)
\sum_n
\phi_n^\dagger(x)\gamma_5\phi_n(x)
\right].
\label{eq:app_Jacobian}
\end{equation}
%\end{widetext}

Using the heat-kernel regulator,
%\begin{widetext}
\begin{equation}
\sum_n\phi_n^\dagger\gamma_5\phi_n
\rightarrow
\lim_{\Lambda\rightarrow\infty}
{\rm tr}
\left[
\gamma_5
e^{-(i\slashed D)^2/\Lambda^2}
\right]_{x,x},
\label{eq:app_heatkernel}
\end{equation}
\end{widetext}
together with
\begin{equation}
(i\slashed D)^2
=
-D^2
-
\frac{g}{2}
\sigma^{\mu\nu}G_{\mu\nu}^aT^a ,
\label{eq:app_Dirac_squared}
\end{equation}
one finds
\begin{equation}
\lim_{\Lambda\rightarrow\infty}
{\rm tr}
\left[
\gamma_5
e^{-(i\slashed D)^2/\Lambda^2}
\right]_{x,x}
=
q(x).
\label{eq:app_heatkernel_result}
\end{equation}
This reproduces the anomalous Ward identity and shows that the anomaly
is a property of the regulated fermion measure
\cite{Fujikawa:1980rc}.

%====================================================

%====================================================

\subsection{Topological susceptibility and axial correlations}

The momentum-dependent topological susceptibility is
\begin{equation}
\chi(q^2)
=
i
\int d^4x\,
e^{iq\cdot x}
\langle0|
T\{q(x)q(0)\}
|0\rangle .
\label{eq:app_chi_q2}
\end{equation}
The flavor-singlet axial-current correlator is
\begin{equation}
\Pi_{\mu\nu}(q)
=
i
\int d^4x\,
e^{iq\cdot x}
\langle0|
T\{J_{5\mu}(x)J_{5\nu}(0)\}
|0\rangle .
\label{eq:app_axial_correlator}
\end{equation}
Lorentz covariance implies the decomposition
\begin{equation}
\Pi_{\mu\nu}(q)
=
(q_\mu q_\nu-q^2g_{\mu\nu})\Pi_T(q^2)
+
q_\mu q_\nu\Pi_L(q^2).
\label{eq:app_correlator_decomposition}
\end{equation}
Taking two divergences and using the anomalous Ward identity in the
chiral limit yields
\begin{equation}
q^\mu q^\nu \Pi_{\mu\nu}(q)
=
(2N_f)^2\chi(q^2),
\label{eq:app_double_divergence}
\end{equation}
and therefore
\begin{equation}
\Pi_L(q^2)
=
\frac{(2N_f)^2}{q^4}\chi(q^2).
\label{eq:app_PiL_chi}
\end{equation}
Thus the longitudinal part of the singlet axial-current correlator is
determined by the topological susceptibility
\cite{Shore:1999be,Shore:2007yn}.
%====================================================

\bibliographystyle{apsrev4-1}
\bibliography{qcd_anomaly_review}

\end{document}